\documentclass[twocolumn,preprintnumbers,amsmath,amssymb]{revtex4}

\usepackage{graphicx}
\usepackage{amsmath,amssymb,amsfonts,amsthm}
\usepackage{dcolumn}
\usepackage{bm}
\usepackage{bbm}
\usepackage{subfigure} 
\usepackage{natbib}
\usepackage[ansinew]{inputenc}
\usepackage{hyperref}

\begin{document}
\title{NONSMOOTH BIFURCATIONS, TRANSIENT HYPERCHAOS AND HYPERCHAOTIC BEATS IN A MEMRISTIVE MURALI-LAKSHMANAN-CHUA CIRCUIT}

\date{\today}
\author{A.~ISHAQ AHAMED}

\address{Department of Physics,Jamal Mohamed College,\\
Tiruchirappalli-620024,Tamilnadu,India}

\author{M.~LAKSHMANAN$^{\dagger}$}

\address{Centre for Nonlinear Dynamics, Bharathidasan University,\\
Tiruchirappalli-620020,Tamilnadu,India\\
lakshman@cnld.bdu.ac.in\\ ~\\
$^{\dagger}$ author for correspondance}~\\~\\

\begin{abstract}
In this paper, a memristive Murali-Lakshmanan-Chua (MLC) circuit is built by replacing the nonlinear element of an ordinary MLC circuit, namely the Chua's diode, with a three segment piecewise linear  active flux controlled memristor. The bistability nature of the memristor introduces two discontinuty boundaries or switching manifolds in the circuit topology. As a result, the circuit becomes a piecewise smooth system of second order. Grazing bifurcations, which are essentially a form of discontinuity induced non-smooth bifurcations, occur at these boundaries and govern the dynamics of the circuit. While the interaction of the memristor aided self oscillations of the circuit and the external sinusoidal forcing result in the phenomenon of beats occurring in the circuit, grazing bifurcations endow them with chaotic and hyper chaotic nature. In addition the circuit admits a codimension-$5$ bifurcation and transient hyper chaos. Grazing bifurcations as well as other behaviors have been analyzed numerically using time series plots, phase portraits, bifurcation diagram, power spectra and Lyapunov spectrum, as well as the recent 0-1 K test for chaos, obtained after constructing a proper Zero Time Discontinuity Map (ZDM) and Poincar\'{e} Discontinuity Map (PDM) analytically. Multisim simulations using a model of piecewise linear memristor have also been used to confirm some of the behaviors. \\ 

\end{abstract}


\maketitle

\section{\label{sec:level1}Introduction} 

The Murali-Lakshmanan-Chua (MLC) circuit which is a two dimensional circuit first introduced by \citet{murali94a}, is basically a classic configuration of forced series LCR oscillatory circuit having a Chua's diode as its non linearity. It is found to exhibit a large variety of bifurcation and chaos phenomena \cite{murali94a,murali94b,murali94c}. An eigen value study of this circuit was made by \citet{lind98}. An exhaustive experimental study on it has been performed by \citet{ml95} and on the MLC and a variant of MLC circuits by \citet{kt00,kt02}. The control and synchronization of chaos in this circuit has been effected by different mechanisms such as nonfeedback methods \citep{murali95,ml96,murali97,raj97} and in the presence of noise by \citet{ram99}. The spatiotemporal dynamics of coupled array of this circuit has been studied by \citet{pm99}. The birth of SNA through type III intermittency route has been reported by \citet{venki99}. While a statistical characterization of the chaotic attractors at bifurcations has been carried out by \citet{phi01}, a new class of chaotic attractors in this circuit has been reported by \citet{ok06}. The observation of chaotic beats in a quasi periodically forced MLC circuit has been reported in \cite{cafag05}. An inductorless realization of this circuit has been designed using current feedback operational amplifiers (CFOA) by \citet{kilic05} and a mixed mode operation of this circuit using CFOA's has been realized by \citet{cam04} and \citet{kilic07}. The presence of multiple period doubling bifurcation route to chaos when periodically pulsed perturbations are applied has been identified by \citet{srini08}. A theoretical study of the memristive MLC circuit has been reported by \citet{wang09a}. Chaos in fractional order MLC circuit has been studied in \cite{wang09b} and in a fractional memristor based MLC system with cubic non linearity by \cite{wang10}. The MLC circuit in the frame of CNN has been studied by \citet{guna10}. The ordered and chaotic dynamics of two coupled MLC like series LCR circuits with a common non linearity has been studied by \citet{santhi11}. Thus we find that over the years the individual, coupled and spatio-temporal dynamics of the MLC circuit have been intensively investigated by a large number of workers. Its importance lies in its low dimensionality, conceptual simplicity as well as the mathematically tractable nature of its nonlinear element - the Chua's diode.

\subsection{Motivation and plan}

The motivation for the present work is the following.
\begin{itemize}
\item
Is it possible to modify the MLC circuit into a memristive MLC circuit by the addition / replacement of its non linearity with a multisim prototype model of the memristor introduced by the present authors \cite{ishaq11} in an earlier work on Chua's circuit?
\item
Will the addition of this memristor make the resultant circuit a non-smooth or piecewise-smooth dynamical system?

\item
If so, what are the types of non-smooth bifurcations that the circuit may admit?

\item
Will the addition of memristor in the MLC circuit result in the same type of switching action and occurrence of beats that were observed in the memristive Chua's circuit?

\item
If so, what will be the nature of the beats: quasiperiodic or chaotic or hyper chaotic?

\end{itemize}

The following work is very much directed towards finding answers to these questions. We have built a memristive MLC circuit, similar to the one reported by \citet{wang09a} numerically, by replacing the Chua's diode with a three segment piecewise flux controlled active memristor deisgned by the same authors earlier, as its non linearity. We have found that the addition of memristor as the nonlinear element converts the system into a second order piece-wise continuous system having two discontinuous boundaries which admit grazing bifurcations. The non-smooth dynamics and the time varying resistive property and the switching characteristic of the memristor give rise to many interesting behaviors such as grazing bifurcations, hyper chaos, transient hyperchaos and hyper chaotic beats in this circuit for a chosen sets of circuit parameters. To our knowledge, we believe that it is for the first time that hyper chaotic beats have been generated in a simple system, non-smooth or otherwise, having just a single driving force. 

The plan of the paper is as follows. In Sec. 2, the characteristic of a Chua-type flux controlled active memristor and its switching properties are discussed and a prototype Multisim model of it, developed by the same authors in an earlier work, is described. In Sec. 3, a modification of the standard MLC circuit by addition of this flux controlled memristor is discussed and the circuit equations are written based upon Kirchoff's laws. The stability analysis of this circuit is also made here. In Sec. 4 a general introduction to non-smooth systems and the types of bifurcations admitted by such systems and the means of observing them correctly are given. In Sec. 5 the classification of the memristive MLC circuit as a nonsmooth system of second order admitting grazing bifurcations is discussed. Further the construction of the Zero Discontinuity Map (ZDM) and the Poincar\'{e} Discontinuity Map (PDM) are also mentioned. In Sec. 6 numerical observation of grazing bifurcation is given. That grazing bifurcations lead to a change in the dynamics of the system has been verified through the application of the 0-1 binary or the K test for chaos, in addition to the usual verification using Lyapunov exponents. The codimension-$5$ bifurcation observed in this circuit is reported in Sec. 7 using phase portraits, Poincar\'{e} maps and bifurcation diagram in the $(\beta-x_3)$ plane. The transient hyperchaos exhibited by this circuit is describried in Sec. 8 and the hyper chaotic beats phenomenon is discussed in Sec. 9. In Sec. 10 a multisim modelling of the memristive MLC circuit is made. With the help of the grapher facility of the software package, time series plots, phase portrait and power spectra  which are qualitatively equivalent to the numerical observations are presented. In Sec. 11 a brief conclusion and possibilities of further studies are discussed.

\section{\label{mem} Memristors and Memristor Characteristic}

A Memristor is a two terminal circuit element supposed to have a functional relationship connecting charge $q$ and flux $\phi$ and endowed with both a resistive property and a memory like nature. This memristor was first postulated by Leon Chua  \cite{chua71} as the missing fourth quartet in the fundamental circuit element family comprising of resistors, inductors and capacitors. Thirty seven years after Chua's prediction, researchers in the HP labs in 2008 revealed its existence by studying the electrical properties of certain nanoscale devices \citep{strut08}. Though no combination of passive devices can reproduce the properties of any fundamental circuit element, much less that of a  memristor, Chua nevertheless theorised that, in principle, memristors with almost any desired $\phi-q$ characteristic can be synthesised using analog circuits \cite{chua71}. Based on this concept, \citet{itoh08} had proposed theoretically a three segment piecewise linear memristor with characteristic curves similar to those of Chua's diode, albeit with a difference. While the characteristic curve of the Chua's diode lies in the ($v-i)$ plane, that of the proposed memristor falls in the ($\phi-q$) plane. Mathematically, this piecewise linear relationship is defined as

\begin{equation}
q(\phi)  =  G_{a_2}\phi + 0.5(G_{a_1}-G_{a_2})[(|\phi+1|)-(|\phi-1|)],
\label{aia:eq1}
\end{equation}
where $G_{a_1}$ and $G_{a_2}$ are the slopes in the inner and outer segments of the characteristic curve. A typical characteristic curve of the Chua type memristor is shown in Fig.  1(a). 

\begin{figure}
\includegraphics[width=08cm]{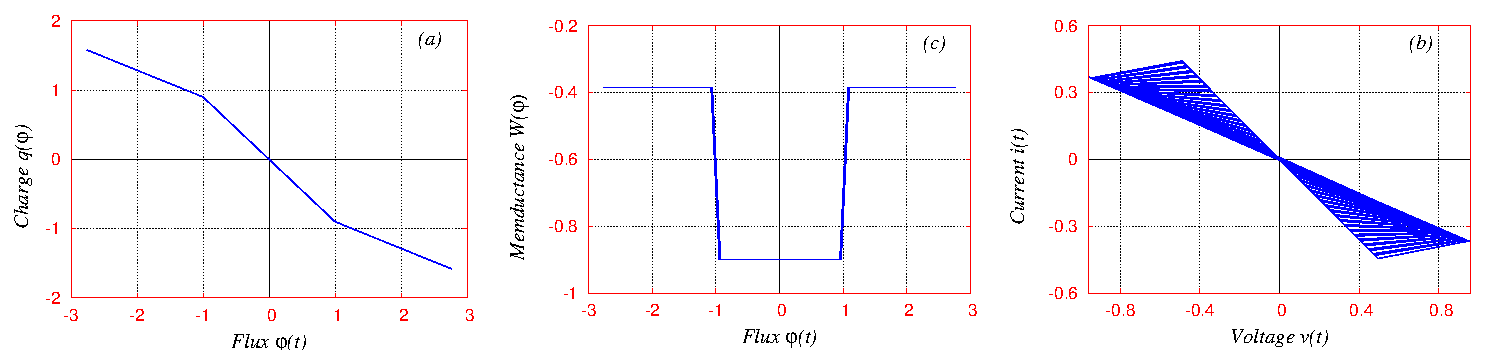}
\caption{ (a) The characteristic curve of the memristor in the $(\phi-q)$ plane, (b)The variation of the memductance $W(\phi)$ as a function of the flux $\phi(t)$ across the memristor  and (c) the variations of current and voltage across the memristor as a result of switching.}
\label{Fig1}
\end{figure}

The current through this Chua-type flux controlled memristor is given by 
\begin{equation}
i(t)  =  W(\phi)v(t),
\label{aia:eq2}
\end{equation}
where $W(\phi)$is the memductance of the memristor. It is so called because it has the units of conductance $(\mho)$ and is given by

\begin{equation}
W(\phi) = \frac{dq(\phi)}{d\phi} = \left\{
					\begin{array}{ll}
					G_{a_1}, ~~~ | \phi  | \leq 1  \\
					G_{a_2}, ~~~ | \phi  | >1
					\end{array}
				\right.
\label{aia:eq3}
\end{equation}
where $G_{a_1},G_{a_2} $ are as defined earlier. 

The memductance of the memristor takes on two particular values as shown in Fig. 1(b), depending on the value of the flux across it. The higher memductance value $G_{a_2}$ can be referred to as the ON state and the lesser memductance value $G_{a_1}$ as the OFF state. Obviously, as the flux across the memristor changes, the memristor switches or toggles between these two states, causing the current and voltage across the memristor to vary as shown in Fig. 1(c). This switching time variant resistive property makes the memristor a desirable element in the modelling of nonlinear circuits. For example, memristor based chaotic circuits and their implementation were described by \citep{muthu09a,muthu09b,muthu10a,muthu10b}. The coexistence of infinitely many stable periodic orbits and stable equilibrium points has been reported in memristive oscillators by \citet{mes10}. Chaos and its control in a four dimensional memristor based circuit using a twin-T Notch filter has been studied in \cite{Iu11}. Transient chaos has been reported in a memristive canonaical Chua's circuit by \citep{bao10a,bao10b}. The dynamical behaviour and stability analysis of a fractional-order memristor based Chua's circuit were made by \citet{pet10}. The nonlinear dynamics of a network of memristor oscillators were investigated by \citet{cor11}. Memristive chaotic circuits based on cellular nonlinear networks were studied by \citet{bus12}.
 
\begin{figure}[h]
\begin{center}
\includegraphics[width=8cm]{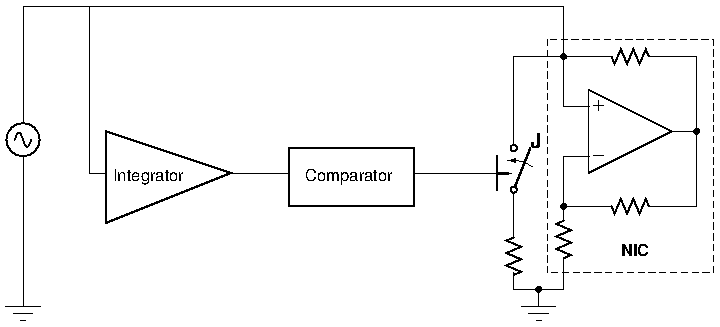}
\includegraphics[width=8cm]{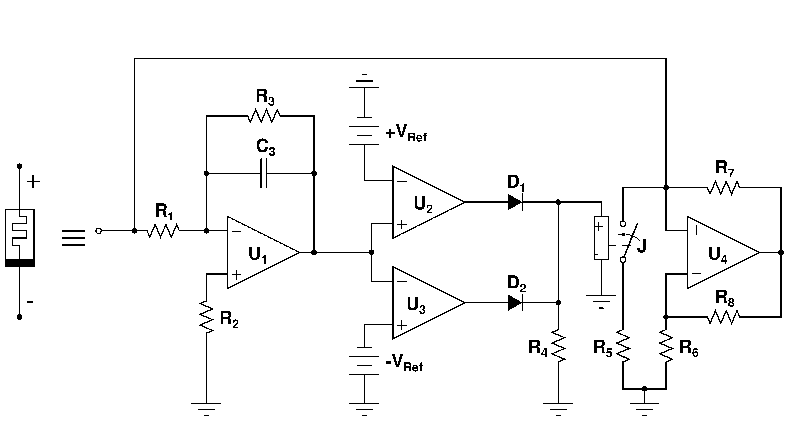}
\end{center}
\caption{ (a) Block diagram for a memristor based on basic principles and (b) a protypical analog circuit model of a three segment piecewise linear flux controlled active memristor used in the study. The parameters are $ R_1 = 10 K\Omega$, $R_2 = 100 K\Omega$, $R_3 = 100 K\Omega$ and $C_3 = 2.2 nF$ for the integrator part, $R_4 = 10 K\Omega$ for the output of the window comparator, $R_5 = 1450 \Omega$ for the linear resistance and $R_6 = 1050 \Omega$, $R_7 = 2 K\Omega$ and $R_8 = 2 K\Omega$ for the negative conductance part. The reference voltages for the window comparator are fixed as $\pm 1$.}
\label{Fig2}
\end{figure} 
 
\subsection{Multisim model of the Memristor}

While different models of the memristor haven been proposed by different authors, see for example, \citet{muthu09b}, we make use of a protypical analog circuit model of a three segment piecewise linear flux controlled active memristor based on Faraday's law $\phi = \int_{0}^{t} vdt$ designed by us, in an earlier work \cite{ishaq11}. The block diagram of this memristor model, is given in Fig. 2(a). It is based on the time varying resistor (TVR) circuit proposed by \citet{nishio93}. Here a linear resistance and a negative impedance converter (NIC) are switched ON and OFF alternatively based on the output pulse of a comparator. The comparator compares the flux through the memristor (that is the integral of the voltage across the memristor) between two reference levels, namely the breakdown points. For flux values lying within the upper and lower breakdown points (namely $\pm 1$ flux units) the negative conductance is included in the circuit. For flux values exceeding the breakdown points, the resultant of the linear resistance and negative conductance, which are in parallel, is included in the circuit. The flux is obtained by an integrator circuit. By this action, the functional relationship between the flux and charge given in equation (1) is realized. A detailed description of this memristor model is given in \cite{ishaq11}.

\begin{figure}[h]
\begin{center}
\includegraphics[width=8cm]{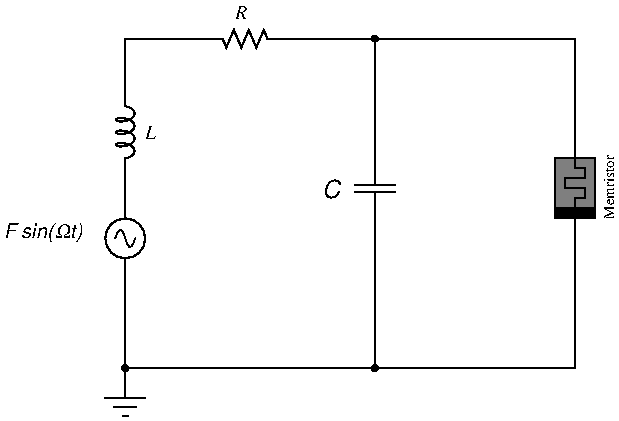}
\includegraphics[width=8cm]{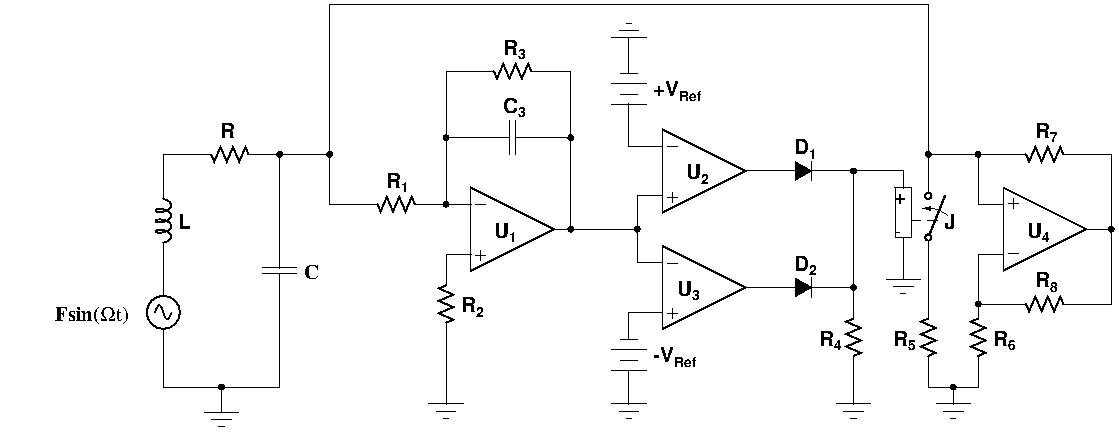}
\end{center}
\caption{(a) The memristive MLC circuit and (b) A Multisim Prototype Model of a memristive MLC circuit. The parameter values of the circuit  are fixed as $ L = 21 mH$, $R = 900 \Omega$, $C = 10.5nF$. The frequency of the external sinusoidal forcing is fixed as $\nu_{ext} = 8.288 kHz$ and the amplitude is fixed as $F = 770 mV ( peak-to-peak voltage V_{pp})$. For the memristive part the parameters are as fixed earlier in Fig. 2. }
\label{Fig3}
\end{figure}

\section{\label{MEM_MLC} Memrestive Murali-Lakshmanan-Chua Circuit}

A Memristive MLC circuit is constructed by replacing the Chua's diode by an active flux controlled memristor as the nonlinear element in the classical Murali-Lakshmanan-Chua circuit. The schematic circuit is shown in Fig. 3(a), while the actual analog realization based on the prototype model for the memristor  is shown in Fig. 3(b). Applying Kirchoff's laws, the circuit equations can be written as a set of autonomous ode's as
\begin{eqnarray}
\frac{d\phi}{dt}  & = & v, \nonumber \\
C\frac{dv}{dt}  & = & i - W(\phi)v,  \nonumber \\ 
L \frac{di}{dt}  & = &  -v -Ri +F\sin( \Omega p),\nonumber \\
\frac{dp}{dt}  & = & 1.
\label{aia:eq4}
\end{eqnarray}
where $W(\phi)$ is the memductance of the memristor and is as defined earlier in Eq. (\ref{aia:eq3}) and $p$ is the time variable in the extended coordinate system.  The normalized form of the circuit equations (7) are
\begin{eqnarray}
\dot{x_1}  & = & x_2, \nonumber \\
\dot{x_2}  & = & x_3-W(x_1)x_2, \nonumber \\ 
\dot{x_3}  & = & -\beta(x_2+x_3) + f \sin(\omega x_4),\nonumber \\
\dot{x_4}  & = &  1.
\label{aia:eq5}
\end{eqnarray}
Here dot stands for differentiation with respect to the normalized time $\tau$ (see below) and $W(x_1)$ is the normalized value of the memductance of the memristor, given as
\begin{equation}
W(x_1) = \frac{dq(x_1)}{dx_1} = \left\{
		\begin{array}{ll}
		a_1, ~~~ | x_1  | \leq 1 \\
		a_2, ~~~ | x_1  | > 1
		\end{array}
	\right.
\label{aia:eq6}
\end{equation}
where $ a_1 = G_{a_1}/G $ and $a_2 = G_{a_2}/G$  are the normalized values of $G_{a_1} $ and $G_{a_2} $  mentioned earlier and are negative. The rescaling parameters used for the normalization are		
\begin{eqnarray}
x_1 = \frac{G\phi}{C},x_2 = v, x_3 = \frac{i}{G}, x_4 = \frac{Gp}{C}, G = \frac{1}{R},\beta = \frac{C}{LG^2}, \\ \nonumber
\omega = \frac{\Omega C}{G} = \frac{2\pi \nu C}{G},\tau = \frac{Gt}{C},f = F\beta.
\label{aia:eq7} 
\end{eqnarray}
\subsection{ Stability Analysis}

When the driving force $f=0$, the system can be considered as a three dimensional autonomous system whose equilibrium state is given by 
\begin{equation}
A = \{ (x_1,x_2,x_3)|\,x_2 = x_3 = 0,\,x_1 = constant\},\nonumber\\ 
\end{equation}
which corresponds to the $x_1$-axis. The Jacobian matrix $D$ at this equilibrium state is given by
\begin{equation}
D_i =   \left ( \begin{array}{ccc}
				0      & 1    	&  0  		 \\
				0      &-a_i    &  1 		 \\
				0      &-\beta  &  -\beta 	\\				 
				\end{array}
		\right).
\label{aia:eq8}
\end{equation}
The characteristic equation associated with the system in this equilibrium state is
\begin{equation}
\sigma^3 + p_2\sigma^2 + p_1 \sigma = 0,
\label{aia:eq9}
\end{equation}
where {\it{$\sigma$}}'s are the eigenvalues and the coefficients $\it{p_i}$'s are given as  $p_2 = (\beta + a_i)$ and $p_1 = \beta(1+a_i)$.
The eigen values that characterize the equilibrium states are given as 
\begin{equation}
\sigma_1 = 0, \,\sigma_{2,3} = \frac{-(\beta+a_i)}{2} \pm \frac{\sqrt{(\beta - a_i)^2-4 \beta}}{2}.
\label{aia:eq10}
\end{equation}
where $ i = 1,2.$
Depending on the eigen values, the nature of the equilibrium states differ. 
\begin{itemize}

\item
When $(\beta - a_i)^2 = 4 \beta$, the equilibrium state will be a stable/unstable star depending on whether $(\beta+a_i)$ is positive or not.

\item
When $(\beta - a_i)^2 > 4 \beta$, the equilibrium state will be a saddle.
muthu10b
\item
When $(\beta - a_i)^2 < 4 \beta$, the equilibrium state will be a stable/unstable focus.
\end{itemize}

In this case the circuit admits self oscillations with natural frequency varying in the range $\frac{\sqrt{(\beta - a_1)^2-4 \beta}}{2} < \omega_o < \frac{\sqrt{(\beta - a_2)^2-4 \beta}}{2}$. It is at this range of frequency that the memristor switching also occurs. Therefore the memristor switching frequency $\omega_{mem}$ depends on the normalized memductance values  $a_1$, $a_2$.

\section{ Non-smooth Systems}

Dynamical systems that contain terms which are non-smooth functions of their arguments are called non-smooth or piecewise-smooth continuous systems. These systems are found to arise everywhere in nature, such as in electrical circuits that have switches, mechanical devices in which components (such as gear assemblies) impact with each other, mechanical systems with sliding, friction, etc. \citep{dib01,dib08}. The common feature of these systems is that they are event driven, in the sense that the smoothness is lost at instantaneous events, for example upon the application of a switch.
The phase space for such a piecewise-smooth system can be divided into two subspaces $S_1 :=\{x\in\mathbb{R}^n, x \leq 0\}$ and $S_2 :=\{x\in\mathbb{R}^n, x > 0\}$ by a discontinuity surface $\Sigma_{12}$ defined by  $\Sigma_{12} = {x\in\mathbb{R}^n:H(x) = 0}$. In each of the subspaces, the dynamics is governed by the smooth vector fields
\begin{equation}
\dot{x}(t) = 
	\begin{cases}
	F_1(x,\mu) & \text{if $x \in S_1$} \\
	F_2(x,\mu) & \text{if $x \in S_2$}.
	\end{cases}
\label{aia:eq11}
\end{equation}
where $x$ denotes state variables. The surface $\Sigma_{12} $ is called the $\it{discontinuity ~set}$ or $\it{discontinuity ~boundary}$ or $\it{switching ~manifold}$.

If $ F_1(x, \mu) \neq F_2(x, \mu)$ at $ x = \Sigma_{12}$, then we have the degree of smoothness to be one. Such systems with the degree of smoothness equal to one are called as $\it{Filippov}$ systems \cite{flip88}. However if $ F_1(x, \mu)=F_2(x, \mu)$ at $ x = \Sigma_{12}$ but there is a difference in the Jacobian derivatives, such that $\frac{\partial F_1(x, \mu)}{\partial x} \neq \frac{\partial F_2(x, \mu)}{\partial x}$ at $ x = \Sigma_{12}$, then the degree of smoothness is said to be two. Systems with degree of smoothness two or above are called $\it{piecewise}$-$\it{smooth ~flows}$. 

Generally as it is impossible to find explicit solutions to dynamical systems, it is convenient to introduce a flow function $\Phi$ generated by the vector field $F(x, \mu)$ such that $\Phi (x_o,t-t_o)$ corresponds to the point at time $t$ on the trajectory that passes through $x_o$ at time $t_o$. Let $\Phi_t$ denote $\frac{\partial\Phi}{\partial t}$, where we have suppressed the parameter dependance for convenience. Then in terms of the flow function, a general dynamical sywhether grazing bifurcations induce a change in the dynamics from periodic to stem can be written as
\begin{equation}
\Phi_t(x,t-t_o) = F(\Phi(x,t-t_o)),\,\Phi(x,0) = x_o,
\label{aia:eq12}
\end{equation}
for all values of $x$ and $t$. Then the unique solution for the dynamical system is given as $x(t)= \Phi(x_o,t-t_o)$. 

When a flow reaches a discontinuity surface, the vector field changes thereby causing a change in the flow function and a sudden qualitative change in the dynamics. These changes arise from a discontinuous jump in the Floquet multipliers and are called as $\it{Discontinuous ~Bifurcations}$ or $\it{ Discontinuity ~Induced ~Bifurcations}$ (DIB)'s \cite{dib08}. A $\it{grazing ~bifurcation}$ is a particular form of discontinuity induced bifurcation which arises when a limit cycle of a flow becomes tangent to ( that is, it just grazes with) a discontinuity boundary. In the following sections we show that the memristive MLC circuit is a second order non-smooth system having two discontinuity boundaries and construct proper discontinuity mappings at these boundaries to observe its dynamics. 

\section{ Memristive MLC Circuit as a Non-smooth System}

The bistability nature of the memristor makes the memristive MLC circuit a piecewise-smooth or non-smooth continuous system of the second order. Referring to the memductance characteristic of the memristor, we find that the memristor switches states at $x = +1$ and at $x = -1$ either from a more conductive ON state to a less conductive OFF state or vice versa depending on how we proceed across the discontinuity boundary. Consequently the phase space can be divided into three subspaces $S_1$, $S_2$ and $S_3$ due to the presence of two switching manifolds $\Sigma_{1,2}$ and $\Sigma_{2,3}$  which are symmetric about the origin. These switching manifolds are defined by the zero sets of the smooth function $H_1(x, \mu) = (x-x^\ast)$, $x^\ast = +1$ and $H_2 = (x-x^\ast)$, $x^\ast = -1$, respectively. The memristive MLC circuit can now be rewritten as a set of smooth ODE's 
\begin{equation}
\dot{x}(t) = 
	\begin{cases}
	F_1(x,\mu) & \text{if $H_1(x, \mu) \geq  0 $ and $H_2(x, \mu) <  0 $ } \\
	F_2(x,\mu) & \text{if $H_1(x, \mu) <  0 $ and $H_2(x, \mu) \geq  0 $ }.
	\end{cases}
\label{aia:eq13}
\end{equation}
where $\mu$ denotes parameter dependance of the vector fields and the scalar functions. The vector fields $F_i$'s are
\begin{equation}
 F_i(x,\mu) =  \left (	\begin{array}{c}
				x_2\\
				x_3-a_i x_2 \\
				-\beta(x_2+x_3)+fsin(\omega x_4)\\	
				1 
				\end{array}
		\right ), \text{i=1,2}.
\label{aia:eq14}
\end{equation}

The conditions for grazing of the periodic orbits (refer Eq. (\ref{aia:eqA.3}) in the Appendix A) are satisfied at both the discontinuity boundaries $\Sigma_{1,2}$ and $\Sigma_{2,3}$. Hence one can check that grazing bifurcations indeed occur in the memristive MLC circuit at both of these discontinuities.
	
Further, the switching action of the memristor induces a tendency in the system to exhibit self-oscillations. Such self-oscillations are known to occur in nonsmooth mechanical systems and were studied extensively by \cite{tys84}. Even the frequencies and amplitudes of such oscillations were estimated using specialized methods. In this system described by Eq. (\ref{aia:eq5}), if an external periodic forcing is applied, there may occur an interaction between this externally induced oscillations and the self-oscillations resulting in the occurrence of beats in the system. This has actually been observed and is described in detail in Sec. 9.

\subsection{Discontinuity Mappings}

Generally non-smooth discontinuity induced birfucations are difficult to analyse in piecewise-smooth systems, because it is necessary to establish the fate of topologically distinct trajectories close to the structurally unstable event that determines the bifurcation. To overcome this difficulty, the concept of the discontinuity map was introduced by \citet{nord91}. This is a synthesised Poincar\'{e} Map whose orbits completely describe the dynamics of the system. Depending on the change in time or otherwise, one can construct two discontinuity maps, namely the Zero-time Discontinuity Map (ZDM) and Poincar\'{e} Discontinuity Map (PDM). For details of deriving anallytically these mappings, one may refer to \citet{dib08}.

If we consider the grazing point as $X_0 = \{x_{10},x_{20},x_{30},x_{40} \} = \{ \pm 1,0,x_{30},x_{40} \}$ then the vector fields of the system become continuous at the grazing point, that is, $F_1(x) = F_2(x) = F(x)$
where \begin{equation}
F(x)  =   \left ( \begin{array}{c}             						
				0      \\						
				x_3  \\						
				-\beta x_3 + f sin(\omega x_4)\\  				
				1   				 				    
				\end{array}
		\right).
\label{aia:eq15}	
\end{equation}
Therefore the system is continuous at the grazing point, but is discontinuous in the first derivative, that is
\begin{equation}
\frac{\partial F_1}{\partial x} \neq \frac{\partial F_2}{\partial x} \nonumber
\end{equation}
This makes the memristive MLC circuit a second order non-smooth system. Following \citet{dib08}, the ZDM and PDM for the memristive MLC circuit can be  derived (for details see Appendix A). The Zero Discontinuity Map is derived as
\begin{equation}
x \mapsto ZDM  = 	\begin{cases}
			x & \text{if $H_{min}(x) \geq  0 $ }\\
			x  + \delta & \text{if $H_{min}(x) < 0$ }.
	\end{cases}
\label{aia:eq16}
\end{equation}
Here 
\begin{equation}
H_{min}(x) = \frac{\partial H}{\partial x}x+O(|x|^2) \nonumber \\
\end{equation}
 where $ O(|x|^2 $ is the correction term in the Taylor expansion of $H(x)$. The ZDM correction $\delta$ is given as
\begin{equation}
\small
\delta  =   \left ( \begin{array}{c}             						
				-\frac{2}{3}\sqrt{2}(a_1-a_2)x_1 \sqrt{\frac{x_1}{x_3}} \\
	                         \frac{2}{3}\sqrt{2}(a_1-a_2)(\frac{x_1}{x_3})^{3/2}{(a_1-2(a_2 + \beta))x_3 + 2fsin(\omega x_4)} \\
				\frac{2}{3}\sqrt{2}(a_1-a_2)\beta x_1 \sqrt{\frac{x_1}{x_3}} \\						
				0				 				    
				\end{array}
		\right).\\
\label{aia:eq17}
\end{equation}
Obviously we find that the correction to the time variable $x_4$ is zero, thereby justifying the nomeclature ZDM. In a similar manner the Poincar\'{e}
Discontinuity Map is derived as
\begin{equation}
x \mapsto PDM  = 	\begin{cases}
			x & \text{if $H(x) \geq  0 $ }\\
			x  + \gamma & \text{if $H(x) < 0$ }.
	\end{cases}	
\label{aia:eq18}
\end{equation}
where the PDM correction $\gamma$ is given as
\begin{equation}
\gamma  =   \left ( \begin{array}{c}             						
				0 \\
	                        \frac{2(a_1-a_2)(a_1+\beta) \sqrt[3]{x_1}}{\sqrt{x_3}}\\
				0\\						
				0				 				    
				\end{array}
		\right).\\
\label{aia:eq19}
\end{equation}
It is essential that these corrections are to be incorporated while deriving the analytical solutions or while performing the numerical integration, so as to obtain a true picture of the non-smooth dynamics. 

\subsection{Event Driven Numerical Simulations}
As the system under consideration is a low dimensional system with just two discontinuity boundaries, discontinuity corrections similar to the above may also be made by following direct numerical simulation using what are known as \emph{event driven}  schemes \cite{dib08}. Often these methods are fast, realistic and accurate.  Here the trajectories in the regions $S_i $ are solved using the standard numerical integration algorithms for smooth dynamical systems such as the Runge-Kutta ODE solver algorithm.  The discontinuity boundary defined as a zero set of a smooth function $H_{ij}(x) = 0 $ is identified and the integration is switched upon crossing the boundary to follow the smooth dynamics pertaining to that particular subspace and appropriate corrections, which are determined by forward and backward integrations, are applied. This scheme therefore reduces the time-integration of a trajectory of a piecewise smooth dynamical system to the problem of finding a set of \emph{event times $t_k$ } and \emph{events} $H_{ij}^{(k)}$ such that 
\begin{equation}
H_{ij}^{(k)}(x(t_k))=0. \nonumber \\
\end{equation}
To implement this one sets up a series of \emph{monitor functions}, the values of which are computed during each integration step. When one of these monitor functions change sign, a root finding method is used to find accurately where $ H_{ij} = 0 $. The corrections are then determined by logical manipulations. In our investigations, we have used both the above methods to validate the concept of discontinuity corrections as applied to the memristive MLC circuit.
\begin{figure}[h]
\begin{center}
\includegraphics[width=8cm]{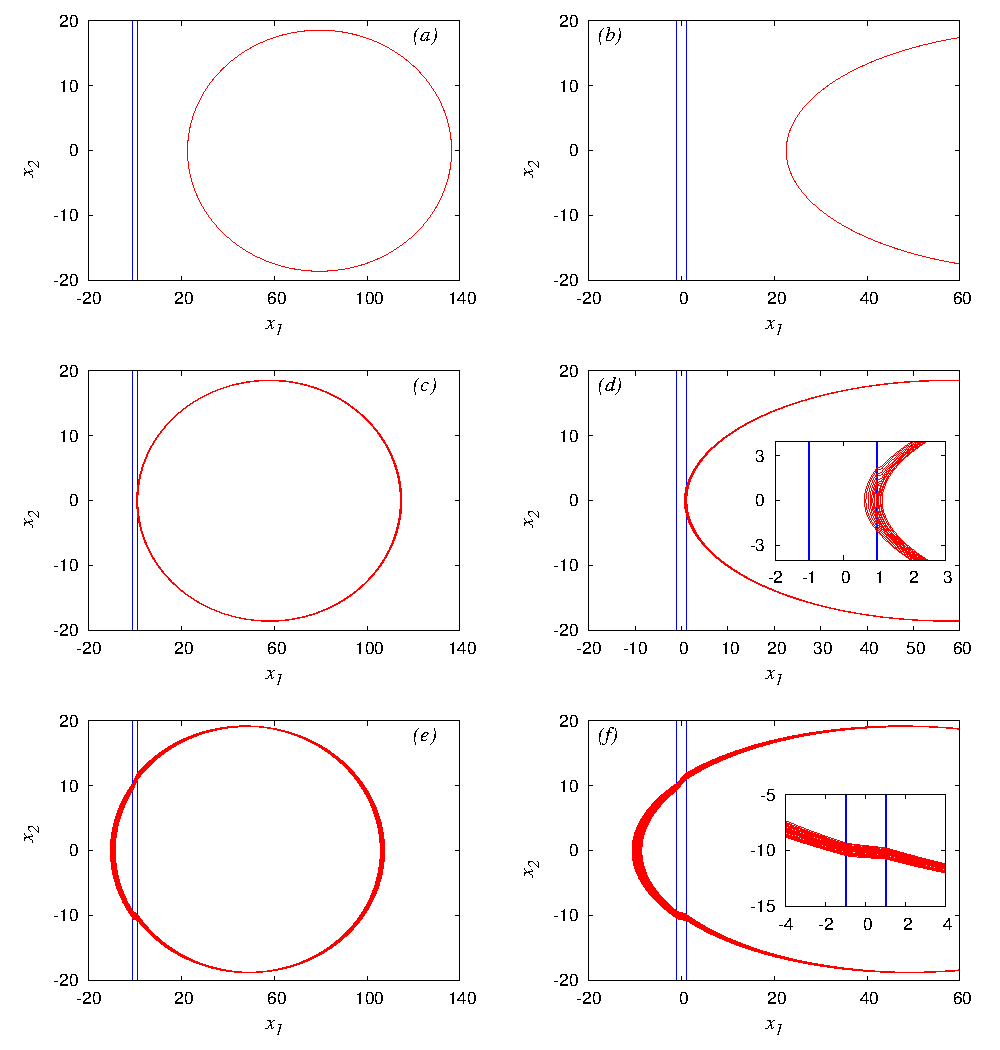}
\end{center}
\caption{Grazing bifurcations for the system defined by Eq. (\ref{aia:eq14}) with ZDM and PDM corrections given by Eqs. (\ref{aia:eq17}) $\&$ (\ref{aia:eq19}) for the parameters $ a_1 = -0.465116$, $a_2 = -0.120289$, $\beta = 0.163613$, $\omega = 0.325940$ and $f = 0.747114$.  A periodic orbit (a) for the initial condition $ \{ 7.0,7.0,7.0,0.0 \}$ as it approaches the grazing boundary $\Sigma = 1 $ flux units in the subspace $S_3$. (c) at the onset of grazing for the initial condition $\{4.5,4.5,4.5,0.0 \}$ and (e) after grazing for the initial condition $\{ 1.5, 2.5,2.5,0.0 \}$. The Lyapunov exponents $\lambda_i's$ in each case are $\{0.000000, -0.022011, -0.022012, 0.000000\}$, $\{ 0.000040,  -0.0218890, -0.021894, 0.000000 \}$ and $\{0.000281,$ $-0.021557, -0.021561, 0.000000 \}$ respectively. Figs. (b),(d) and (f) are the blown up portions of the orbit in Figs. (a), (c) and (e) respectively. The insets in Figs. (d) $\& $ (f) show the trajectories at grazing and after complete onset of grazing with more clarity. }
\label{Fig4}
\end{figure}

\section{ Grazing Bifurcations }

In the presence of forcing, we find that the circuit admits limit cycle motion resulting in periodic orbits. When a periodic orbit impacts the discontinuity surfaces tangentially,  a local stretching of the phase space leading to a profound destabilizing effect on the dynamics have been observed. We find that the periodic orbit after impact is thrown away from the discontinuity surfaces to regions infinitely deep in the subspaces $S_1$ or $S_3$. Hence recording this grazing bifurcation numerically is difficult. To overcome this problem, the parameters are fixed at some desired values and the initial conditions are changed starting from faraway regions in subspaces $S_1$ or $S_3$, to regions close to the discontinuity boundaries. By following this approach it is possible to find the behaviour of the periodic orbit in phase space, before impact, just at the instant of impact and after the impact has taken place with the discontinuity boundary. The parameter values are fixed as $a_1 = -0.465116$, $a_2 = -0.120289$, $\beta = 0.163613$, $ \omega = 0.325940$ and $f = 0.747114$. For an initial condition $ X_0 =  \{ 7.0,7.0,7.0,0.0 \} $ we find a periodic orbit close to the discontinuity boundary. This is shown in Fig. 4(a). A blown up portion of a section of the periodic orbit is shown in Fig. 4(b) for clarity. The Lyapunov exponents for this periodic orbit are $\lambda_1 = 0.000000, \lambda_2 = -0.022022,\lambda_3 = -0.022012$ and $\lambda_4 = 0.000000$. As the initial point is brought closer to the discontinuity boundary, grazing sets in. This is shown in Fig. 4(c) $\&$ 4(d) for the initial condition $X_0 = \{ 4.25,4.25,4.25,0.0 \}$. The corresponding Lyapunov exponents are $\lambda_1 = 0.000040, \lambda_2 = -0.021890,\lambda_3 = -0.021894$ and $\lambda_4 = 0.000000$. For an initial condition $ X_0 = \{1.5,2.5,2.5,0.0 \}$ we find that the periodic orbit has been blown up into a fully chaotic attractor with $\lambda_1 = 0.000281, \lambda_2 = -0.021557,\lambda_3 = -0.021561$ and $\lambda_4 = 0.000000$. These are shown in Figs. 4(e) $\&$ 4(f). Thus we find that the attractor which is purely periodic before transversal intersection with the discontinuity boundary, becomes chaotic as a result of grazing bifurcations. This result is also proved by applying the K-test for chaos as described in the next section.

\subsection{\label{K-test} 0-1 Test for Chaos}

Recently \citet{got04} have introduced a new test called the binary test or the 0-1 test or the K-test for chaos. This test uses a scalar time series data obtained from the system to derive a rotation coordinate $\theta$ and translation coordinates $(p,q)$ to form a Euclidean group $E(2)$. As reported by \citet{nicol01}, the group extensions of this Euclidean group are sublinear and are 
\begin{itemize}
\item
bounded when the system dynamics is periodic or quasiperiodic and
\item
unbounded random Brownian motion like behaviour when the system dynamics is chaotic
\end{itemize}

This causes the mean square displacement $K$ of points in the space of the translation variables, namely the  $(p-q)$ space, to become $0$ when the system behaviour is periodic and $1$ if it is chaotic. 
The main advantages of this test are that 
\begin{itemize}
\item
the dimension of the dynamical system and the form of the underlying equations are not important
\item
the determination of embedding dimension and delay time, the essential requisites of phase space 
reconstruction, are done away with
\item
there is an element of definiteness. This is because a value of 0.01 for K denotes definitely a nonchaoic 
behaviour, which is not possible with the lyapunov exponents
\item
it provides for a straight forward visualisation of the dynamics in the translation variables $(p-q)$ space.
\end{itemize}

The success of this procedure depends on leaving long transients such that the system behaviour settles down 
to its asymptotic state and having larger values of integration time $N$. 
The test is universally applicable to any deterministic dynamical system, in particular to ordinary and 
partial differential equations, and to maps \cite{got04}, as well as to deterministic systems with noise \cite
{got05}. This test has also been applied successfully to Chua's circuit\cite{aziz12}, to fractional Chua's 
citcuit \citep{grassi08,grassi10} and to fractional simplified Lorenz oscillator \cite{Ke10}.

Due to the advantages mentioned above, the 0-1 test is getting greater acceptance by nonlinear researchers 
for characterizing the dynamics of nonlinear systems. 

\subsection{\label{Appln of K-test}  Confirmation of Grazing Bifurcation Induced Dynamics using 0-1 Test for Choas}

The 0-1 test can be applied to the memristive MLC circuit described by Eq. (14) along with the ZDM and PDM 
corrections given by Eqs. (17) $\&$ (19), to find out whether grazing bifurcations induce a change in the 
dynamics from periodic to chaotic behaviour, independent of the values of the associated Lyapunov exponents.  
To apply this test, the translation coordinates $p(t) \& q(t)$ given by \cite{got04} are obtained as

\begin{eqnarray}
p_c(n) & = & \sum_{j=1}^{n}\phi(j)\cos(\theta_c), \nonumber \\ 
q_c(n) & = &  \sum_{j=1}^{n}\phi(j)\sin(\theta_c), \nonumber \\ 
\theta_c(j) & = &  jc + \sum_{i=1}^{j}\phi(i).
\label{aia:eq20}
\end{eqnarray}
where  $\phi(j)$ is the scalar time series for any variable of the system of length $ j = 1,2,3.......N $, $c \in (0,\pi)$. The mean square displacement of the translation variables is given for a given value of $c$ as
\begin{equation}
\small
M_c(n) = \lim_{N \rightarrow \infty } \frac{1}{N-n}\sum_{j=1}^{N-n}[p_c(j+n)+ p_c(j)]^2 + [q_c(j+n)+ q_c(j)]^2
\label{aia:eq21}
\end{equation}
where $ n = 0,1,2..........N/10 $. The above mean square displacement grows linearly for chaotic behaviour 
but remains bounded for nonchaotic behaviour. The asymptotic growth rate of this quantity can be found from 
linear regression as
\begin{equation}
K_c = \lim_{n \rightarrow \infty }\frac{log M_c(n)}{log(n)}
\label{aia:eq22}
\end{equation}
The median of $K_c$ for $c \in (0,\pi)$ denoted as $K$ gives the desired characterization parameter. 

The attractors of the memristive MLC circuit in the space of the translation variables, namely the $(p-q)$ space for the periodic behaviour before grazing, for the behaviour at grazing and for the chaotic behaviour after grazing are shown in Figs. 5(a), 5(c) $\&$ 5(e) respectively. The corresponding variations in the values of the mean square displacement $K_c$ as a function of the constant $c$ are shown in Figs. 5(b), 5(d) $\&$ 5(f). The median $K$ of these $K_c$ values is obtained for each case and are given as  $K = 0.000440$ for the system's behaviour before grazing and  $K = 0.000661$ for behaviour at grazing and $K = 0.998337$ for the system's behaviour after grazing. These values of $K$ prove that grazing bifurcations occurring in the system are indeed responsible for the change in the dynamics of the system from a periodic behaviour to chaotic behaviour.

\begin{figure}[h]
\begin{center}
\includegraphics[width=8cm]{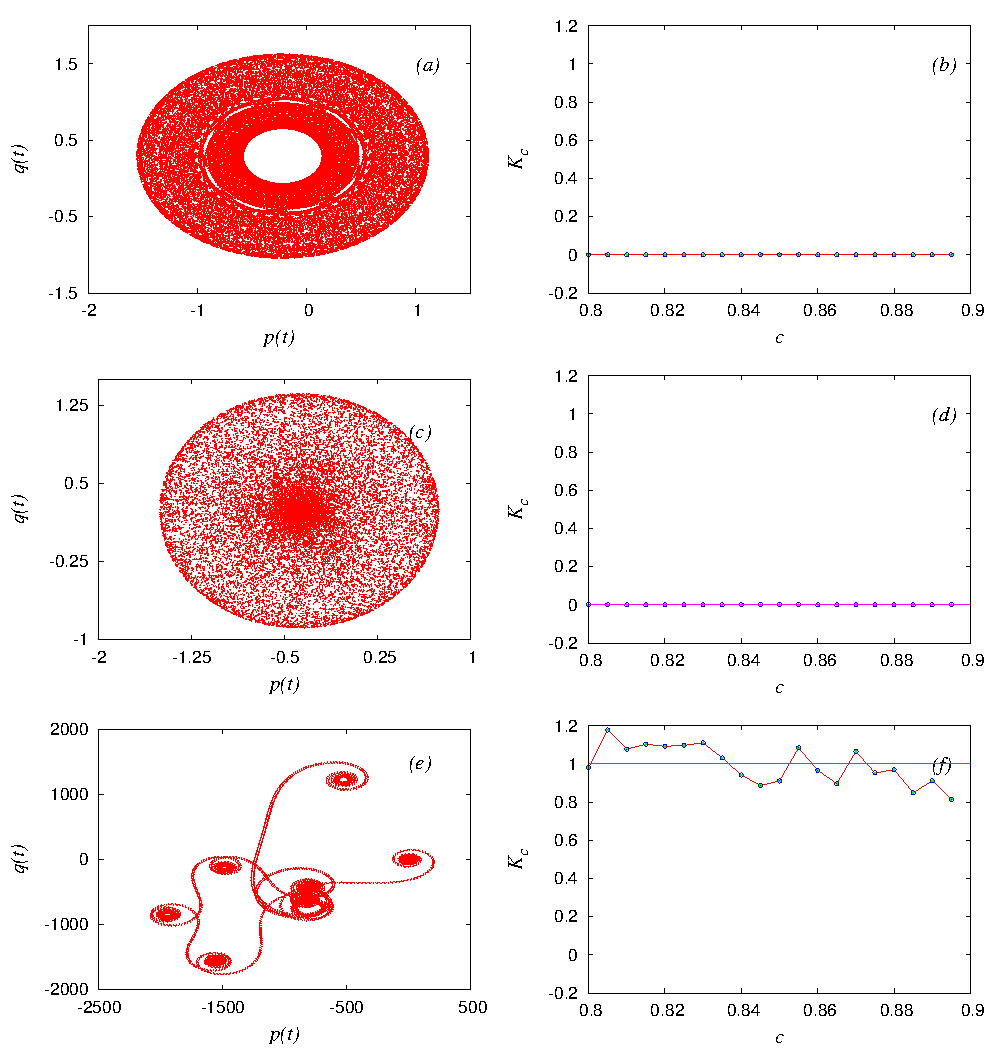}
\end{center}
\caption{ Attractors of the system defined by Eq. (\ref{aia:eq14}) in the plane of the translation variables, namely the $(p-q)$ plane for (a) periodic behaviour before grazing , (c) for the periodic behaviour at grazing and (e) for the chaotic behaviour after grazing. (b),(d) $\&$ (f) the corresponding variations in the mean square displacement K values  as a function of the constant  $c$. }
\label{Fig5}
\end{figure}

\section{\label{Co-dim} Codimension-\emph{m} Bifurcation }

A codimension-\emph{m} bifurcation refers to a bifurcation scenario induced by the concurrent variation of $m
$ parameters of the system. To unfold this bifurcation fully one requies a $m-dimensional$ parameter space \cite{nayfeh95}. Let us consider the case where the the values of the circuit elements are fixed as $ C = 12.5 nF$, $L = 19.1mH$, $\Omega = 2\pi\nu$, where $\nu = 8300Hz$ is the frequency and $F = 1000$ mVpp the peak-to-peak value of the external force. As the series resistance $R$ enters into the expressions of all the 
normalized parameters given in Eq. (7), it is but natural that all of them change when the resistance alone 
is varied. 

\begin{figure}[h]
\begin{center}
\includegraphics[width=8cm]{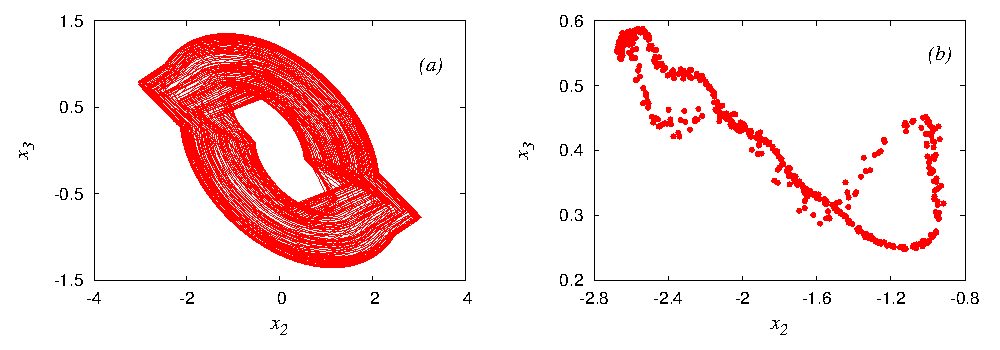}
\end{center}
\caption{(a) The phase portrait of the hyperchaotic attractor in the $(x_2 - x_3)$ plane and (b) its corresponding Poincar\'{e} map. The series resistance for this case is $R= 720 \Omega$ giving the normalized parameter values as  $a_1 = -0.669767$, $a_2 = -0.173216$, $\beta = 0.339267$, $\omega =   0.469354$, $f = 0.119949$ and $h =  0.013387$.}
\label{Fig6}
\end{figure}

For example, when $R = 720 \Omega$, the normalized parameters of the system are $a_1 = -0.669767$, $a_2 = 
-0.173216$, $\beta = 0.339267$, $\omega =  0.469354$, $f = 0.119949$ and $h =  0.013387$. For this choice of 
parameters, the behaviour is hyper chaotic with the associated Lyapunov exponent values as $( \lambda_1 = 
0.136691,\lambda_2 = 0.034005,\lambda_3 = -0.109690,\lambda_4 = 0.000000)$. The phase portrait of this 
hyperchaotic attractor in the $(x_2-x_3)$ plane and its corresponding Poincar\'{e} map are shown in Figs. 6(a)
$\&$ (b).

\begin{figure}[h]
\begin{center}
\includegraphics[width=8cm]{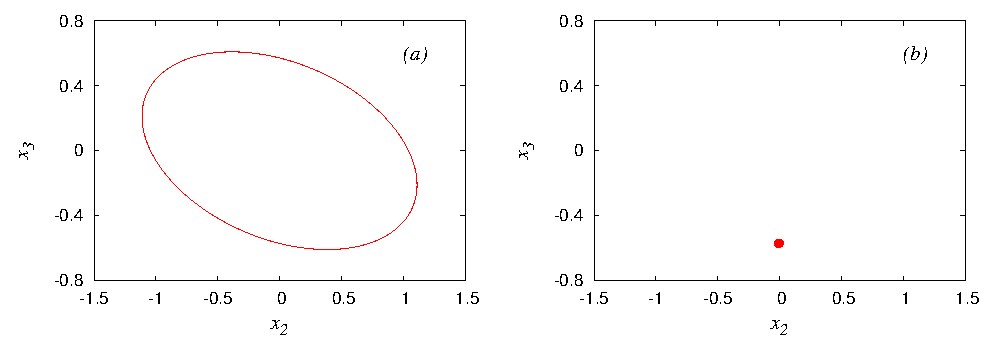}
\end{center}
\caption{ (a) The phase portrait of the periodic attractor in the $(x_2 - x_3)$ plane and (b) its corresponding Poincar\'{e} map. The series resistance for this case is $R = 788 \Omega$ causing the normalized parameters to be $a_1 = -0.733002$, $a_2 = -0.189577$, $ \beta =  0.406377$, $\omega =   0.513682$, $f =  0.143676$ and $h =  0.012232$.}
\label{Fig7}
\end{figure}

However when $R =788 \Omega$, the normalized parameters get changed as $a_1 = -0.733002$, $a_2 = -0.189577$, $ \beta =  0.406377$, $\omega =   0.513682$, $f =  0.143676$ and $h =  0.012232$. The phase portrait of the system for this choice of parameters in the $(x_2-x_3)$ plane and its corresponding Poincar\'{e} map are shown in Figs. 7(a)$\&$ (b). The values of the associated Lyapunov exponents are $( \lambda_1 = -0.000052, \lambda_2 = -0.116821,\lambda_3 = -0.117559,\lambda_4 = 0.000000)$. 

The power spectrum for the system shown in Fig. 8(a) when the series resistance $R = 720 \Omega$ is of broadband nature, indicating the hyperchaotic nature of the attractor. However the power spectrum in Fig. 8(b) for the case when $R = 788 \Omega$, contains a single peak, which clearly reveals the behaviour of the system to be periodic. 

\begin{figure}[h]
\begin{center}
\includegraphics[width=8cm]{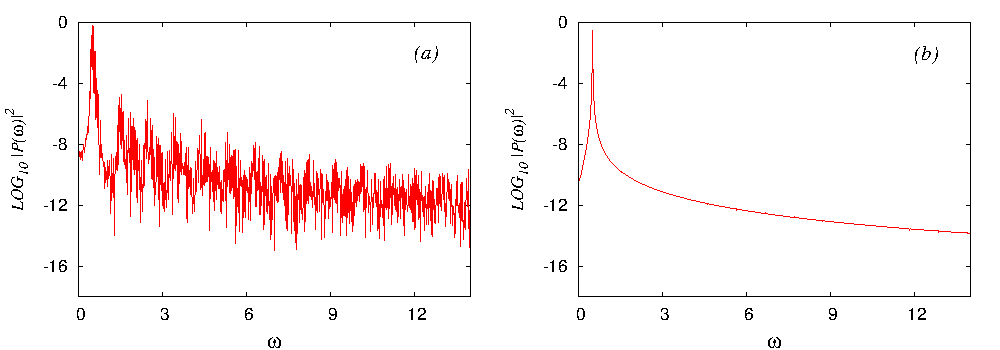}
\end{center}
\caption{The power spectrum of (a) the hyperchaotic attractor and (b) the periodic attractor. }
\label{Fig8}
\end{figure}

\begin{figure}[h]
\begin{center}
\includegraphics[width=8cm]{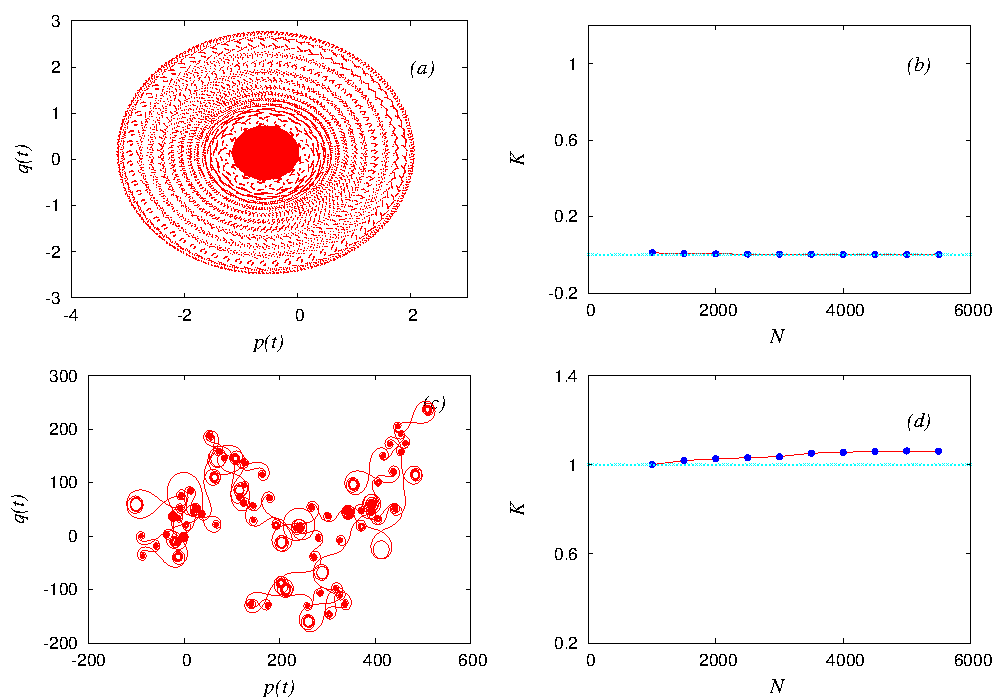}
\end{center}
\caption{ The phase portraits of (a) the periodic attractor and (c) the hyperchaotic attractor in the plane of the translational variables, namely the $(p-q)$ plane. (b) $\&$ (d) The variation of the $K$ values as a function of discrete time $N$ for these two cases respectively.}
\label{Fig9}
\end{figure}

Further the hyperchaotic and the periodic natures of the system's behaviour are also confirmed by the phase portraits in the plane of the translational variables, namely the $(p-q)$ plane as shown by Figs. 9(a) $\&$ (c) and in the variation of the $K$ values shown in Figs. 9(b) $\&$ (d), for a constant value of the random constant $c$ in Eq(\ref{aia:eq20}), for a range of values of the discrete time $N$. It has been found that the median of these $K$ values for the hyperchaotic attractor is $K = 1.04475$ and for the periodic case $K = 0.003138$. 
 
The stable hyperchaos and periodicity exhibited by the system is a consequence of the nonsmooth grazing bifurcation undergone by the system. 

\subsection{\label{bif_diagram} Bifurcation Diagram}

As it is difficult to depict this codimension-\emph{5} bifurcation in three dimensional space, we capture this bifurcation with the help of a bifurcation diagram in a two dimensional plane. For this purpose we vary the series resistance in the system in Fig. 3 in the range $ 400\Omega < R < 1000\Omega $, starting from $1000\Omega$. Due to this variation in resistance, the normalized parameters vary as follows, $ -0.930233 < a_1 <  -0.372093$, $ -0.240577 < a_2 <  -0.096231 $, $ 0.104712 < \beta <  0.654450$, $  0.260752 < \omega_{ext} < 0.651880$ and $  0.037021 < f <  0.231383 $. These variations in parameters cause the system to transit from a stable periodic attractor to a hyper chaotic attractor suddenly at a resistance value $ R= 740 \Omega $ or equivently $\beta =  0.358377$ . Hence we can say that the system undergoes a codimension-\emph{5} bifurcation. The circuit then alternates between hyper chaotic, chaotic and transient hyper chaotic states as shown in the in the $(\beta - x_1)$ plane representation of the said bifurcation in Fig. 10(a). The transient hyper chaotic states are shown by blank regions or gaps in the bifurcation diagram. This is because at the instants of transition from hyper chaotic to periodic state, the attractor is thrown from the regions in subspaces $S_1$, $S_2$ and $S_3$ close to the fixed point at the origin to infinitely faraway regions in either the subspace $S_1$ or $S_3$, where it eventually settles into a chaotic state. The Lyapunov spectrum for the same range as in the bifurcation diagram is shown in Fig. 10(b). This spectrum follows closely the dynamics represented by the bifurcation diagram.

\begin{figure}[h]
\begin{center}
\includegraphics[width=8cm]{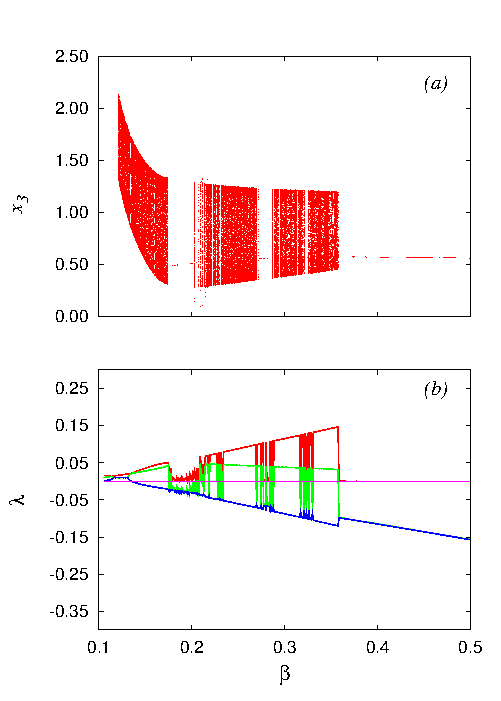}
\end{center}
\caption{(a) The bifurcation diagram in the $(\beta-x_3)$ plane corresponding to the codimension-5 bifurcation and (b) the corresponding Lyapunov spectrum. }
\label{Fig10}
\end{figure}

In order to ensure that our numerical simulation follows closely the dynamics, we have taken the step size as  $h = (2\pi)/(\omega N)$, where we have chosen $N = 1000$. This gives the value of the step size as $h = 0.0193$. Also as the system represented by Eq. (5) is a non-autonomous system, this choice of step size is not altogether new, as this choice has to be necessarily made if one wishes to take a stroboscopic section of the attractor, see for example \cite{park86}. However the dynamics is independent of the step size of the numerical integration. To verify this, we have also repeated our simulations with  a smaller step size, by choosing $N = 1500$ and $N = 2000$ and have found the dynamics to remain the same. Also the results obtained using these step sizes are similar to those one may get using the standard value for the step size, namely $h = 0.01$. 

\begin{figure}[h]
\begin{center}
\includegraphics[width=8cm]{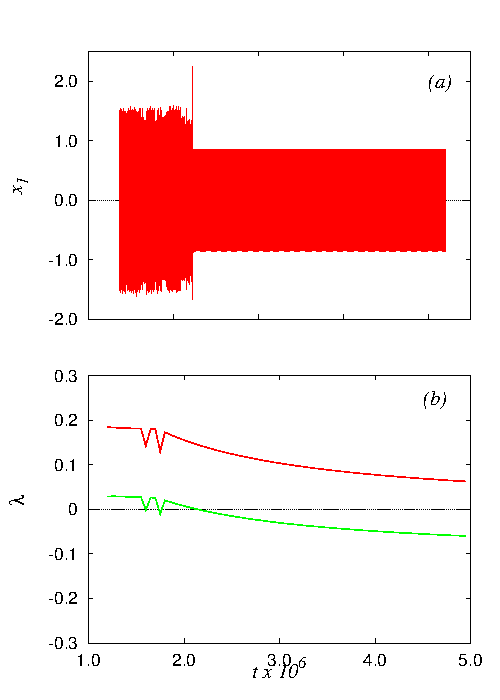}
\end{center}
\caption{(a) The time series plot showing transient hyperchaos and (b) the Local Lyapunov spectrum confirming the same.The parameters cosidered are $a_1 = -0.725581$, $a_2 = -0.187650$, $\beta = 0.398168$, $\omega_{ext} = 0.505404$, $f=0.200602$ and $h=0.012432$. }
\label{Fig11}
\end{figure}

\begin{figure}[h]
\begin{center}
\includegraphics[width=8cm]{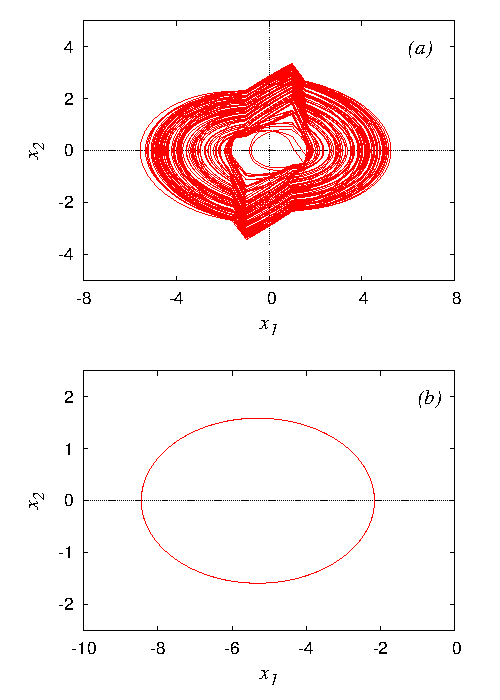}
\end{center}
\caption{The phase portrait in the $(x_1 - x_2)$ plane for (a) the transient hyperchaotic state upto approximately $2.2$ x $10^6$ normalized time units and (b) in the asymptotic state beyond $2.2$ x $10^6$ normalized time units. }
\label{Fig12}
\end{figure}

\section{\label{TH}Transient Hyperchaos}

Transient chaos is a ubiquitous phenomenon observed in many nonlinear dynamical systems, wherein a trajectory behaves chaotically for a finite amount of time before settling into a final (usually nonchaotic) state. This arises due to the presence of nonattracting chaotic saddles in phase space 
\citep{greb82,greb83,kantz85,hsu88,tel90,tel96}. It is known that chaotic saddles and transient chaos are responsible for important physical phenomena such as chaotic scattering \cite{ott93}, and particle transport in open hydrodynamical flows \cite{pen95}. They are belived to be the culprit for catastrophic phenomena such as voltage collapse in electrical power systems \citep{dob89,wang92,mukesh99} and species extinction in ecology \cite{cann94}. An extensive study of chaotic transients in spatially extended systems was studied by \citet{tel08}. It has also been reported in a memristive Chua oscillator by \citep{bao10a,bao10b}. In our present study we have found transient hyperchaos in the memristive MLC circuit. To observe this we set the parameters of the circuit as $a_1 = -0.725581,\,a_2 = -0.187650 $ and $ \beta=0.398168 $. The frequency of the external force is set as $\omega_{ext} = 0.505404$ and the strength of the force as $f = 0.200602$. The step size for the numerical integration is assumed as $ h = 0.012432$. The initial conditions are chosen as $ X_0 = \{1.0,-0.1,0.1,0 \}$. For these values of the parameters, the circuit initially exhibits hyperchaos and then switches over to chaotic oscillations. This can be seen in the time series plot shown in Fig. 11(a). The local Lyapunov spectrum, drawn as a function of time shown in Fig. 11(b), illustrates clearly this transition. While the two largest local Lyapunov exponents (LLE) are positive initially, at the instant of the transition to chaos, the second largest local exponent changes from positive to negative values. The phase portraits of the oscillator in the $x_1-x_2$ plane in the transient hyper chaotic and asymptotically chaotic states are shown in Figs. 12(a) $\&$ 12(b). Though the attractor shown in Fig. 12(b) appears as a period 1-T orbit, it is in fact an overlay of different trajectories. As one tries to compute the solution with greater accuracies, the attractor continues to fill out, thus justifying the low positive values of the largest local Lyapunov exponent in the asymptotic state. This is a consequence of the non-smooth nature of the system.

\begin{figure}[h]
\begin{center}
\includegraphics[width=8cm]{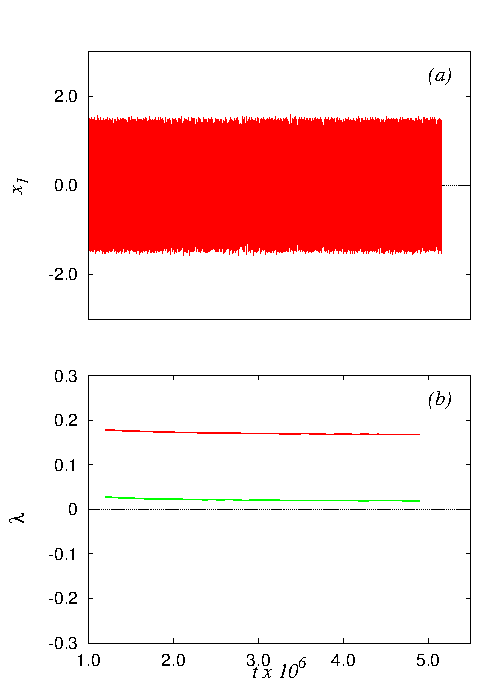}
\end{center}
\caption{(a) The time series plot showing persistent or stable hyperchaos for a slightly different set of parameters $a_1 = -0.725581$, $a_2 = -0.187650$, $\beta = 0.398168$, $\omega_{ext} = 0.508467$, $f=0.183006 $ and $ h=0.012357$ and (b) the Local Lyapunov spectrum confirming the same.}
\label{Fig13}
\end{figure}

\begin{figure}[h]
\begin{center}
\includegraphics[width=8cm]{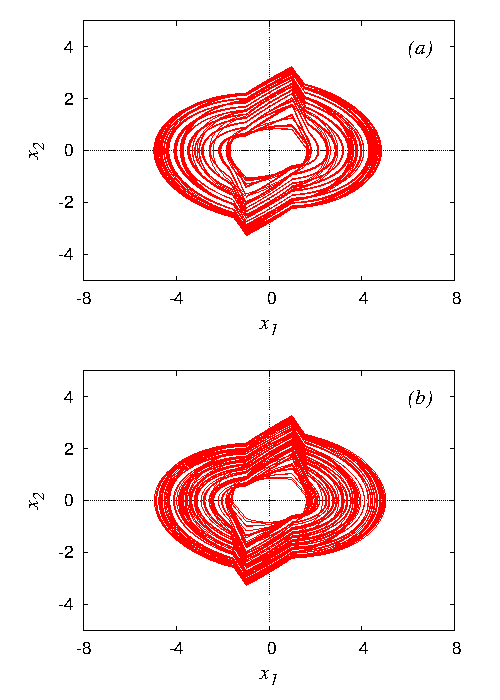}
\end{center}
\caption{The phase portrait in the $(x_1 - x_2)$ plane for (a) the persistent hyperchaotic state at approximately $2.2$ x $10^6$ normalized time units and (b) in the asymptotic region beyond $2.2$ x $10^6$ normalized time units.}
\label{Fig14}
\end{figure}

In contrast to the transient hyperchaos, persistent or stable hyperchaos observed for slightly different parameter values $a_1 = -0.725581,\,a_2 = -0.187650 $ and $ \beta=0.398168 $, $\omega_{ext} = 0.508467$,$f = 0.183006$ and $ h = 0.012432$ is shown by the corresponding time series, local Lyapunov spectrum in Figs. 13(a) $\&$ 13(b) and phase portraits in Figs. 14(a) $\&$ 14(b).

\section{\label{HB}Hyperchaotic Beats}

Beats is a phenomenon arising out of the interference between two different oscillations having small difference in frequencies. This phenomenon has been widely studied in linear systems. However chaotic and hyper-chaotic beats are found to occur in nonlinear systems also. Chaotic and hyper-chaotic beats were identified for the first time in a system of coupled Kerr oscillators and coupled Duffing oscillators with small nonlinearities and strong external pumping \cite{grygiel02}. Chaotic beats were also reported in coupled non-autonomous Chua's circuits \cite{cafag04}. Weakly chaotic and hyper-chaotic beats were reported in individual and coupled nonlinear optical subsystems, respectively, describing second harmonic generation (SHG) of light by \citet{sliwa08}. In all these coupled systems \citep{grygiel02,cafag04,sliwa08} the occurrence of beats is explained as due to the interaction between the self oscillations or driven oscillations of each of the coupled subsystems. Using extensive Pspice simulations, the occurrence of beats in individual driven Chua's circuit with two external excitations has been reported \citep{cafag06a,cafag06b}. Chaotic beats have also been found to occur in individual Murali- Lakshmanan-Chua (MLC) circuit by the same authors \citep{cafag05}. In these cases of single systems \citep{cafag05,cafag06a,cafag06b}, two different external driving forces with slightly differing frequencies were the cause for the occurrence of beats. Chaotic beats in a Memristive Chua's circuit also was reported earlier by \cite{ishaq11}. In the present work, we have identified hyper chaotic beats in the memristive MLC circuit. The phase portrait in the $(x_1-x_2)$ plane, the chaotic switching of the memrsitor are shown in Figs. 15(a-b). The power spectra of $x_1$ variable as well as that of the memristor switching are shown in Figs. 15(c-d), respectively. The time series of the modulated normalized flux $x_1$ is shown for an extended range in Fig. 16(a),  while a section of the same is shown in Fig. 16(b). Similar results obtained by numerical simulation employing $\emph{even-driven}$ scheme are shown in Figs. (17 $\&$ 18).

\begin{figure}[h]
\begin{center}
\includegraphics[width=8cm]{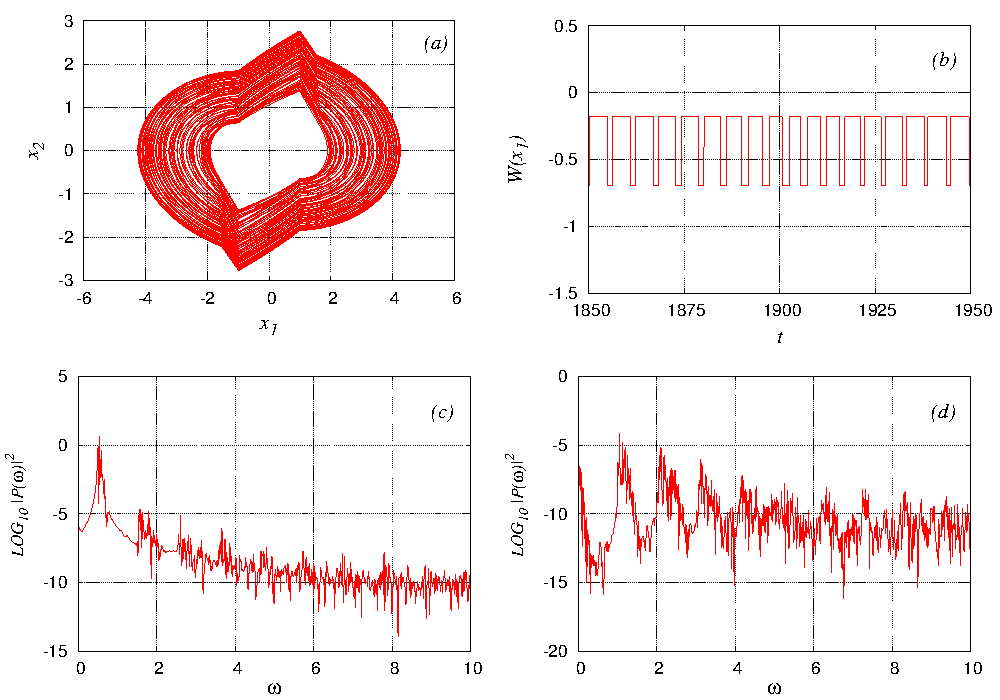}
\end{center}
\caption{ \emph {Hyperchaotic Beats: Numerical Simulations with Analytical Corrections} : The phase portrait in (a) the $(x_1-x_2)$ plane, (b) the variation in time of the memductance $W(x_1)$, (c) the power spectrum of the normalized flux $x_1$ across the memristor and (d) the power spectrum of memristor switching showing it to be chaotically time varying resistor (CTVR). The parameters are $\beta = 0.355263$, $a_1 = -0.697674$, $a_2 = -0.180433$, $\omega_{ext} = 0.508938$ and $f = 0.081643$. The Lyapunov exponents $\lambda_i's$ are $\{0.125510, 0.016904, -0.017554, 0.000000 \}$. }
\label{Fig15}
\end{figure}

\begin{figure}[h]
\begin{center}
\includegraphics[width=8cm]{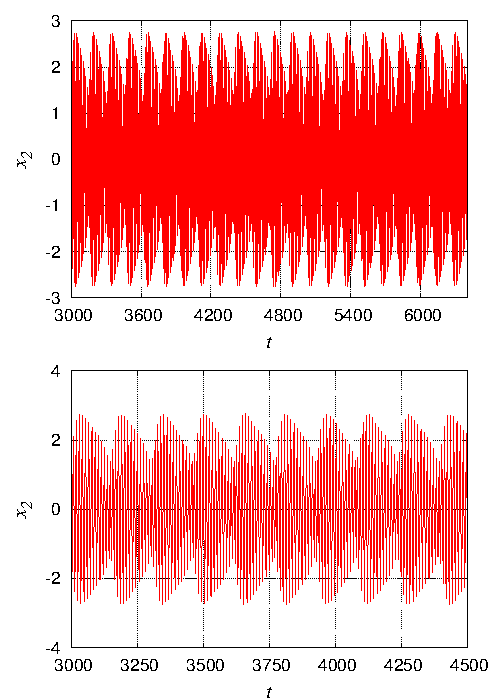}
\end{center}
\caption{\emph {Hyperchaotic Beats: Numerical Simulations with Analytical Corrections} : (a)The time series of the hyperchaotically modulated $x_2$ variable and (b) an expanded portion of the same for a short stretch of time.}
\label{Fig16}
\end{figure}

\begin{figure}[h]
\begin{center}
\includegraphics[width=8cm]{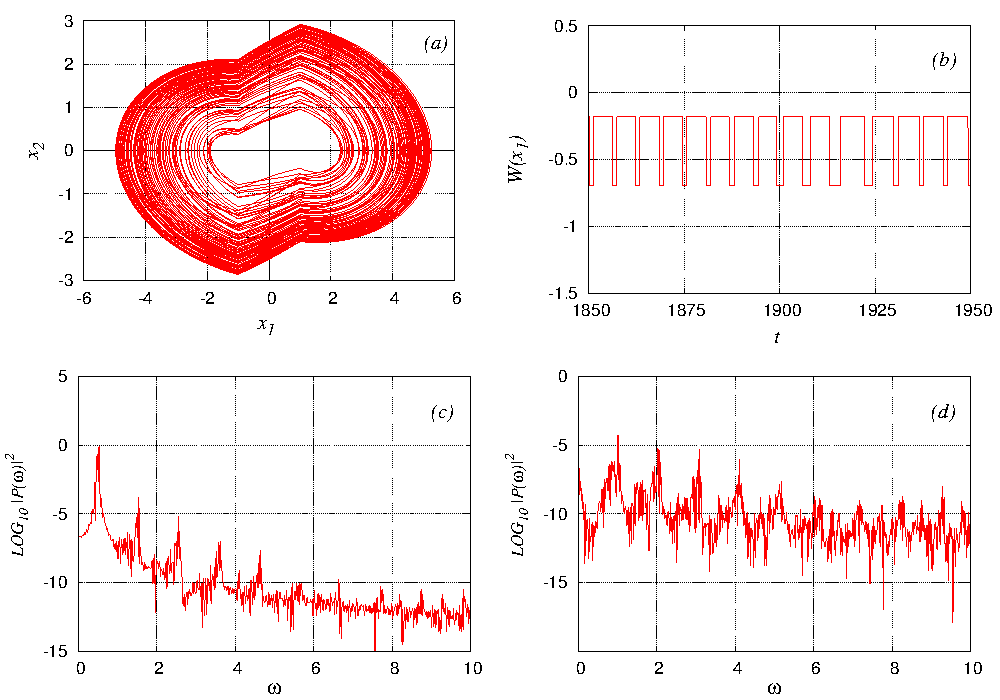}
\end{center}
\caption{\emph {Hyperchaotic Beats: Numerical Simulations (Event Driven Scheme)} : The phase portrait in (a) the $(x_1-x_2)$ plane, (b) the variation in time of the memductance $W(x_1)$, (c) the power spectrum of the normalized flux $x_1$ across the memristor and (d) the power spectrum of memristor switching showing it to be chaotically time varying resistor (CTVR). The parameters are $\beta = 0.355263$, $a_1 = -0.697674$, $a_2 = -0.180433$, $\omega_{ext} = 0.528730 $, $f = 0.081643$ and h = 0.011884.}
\label{Fig17}
\end{figure}
\begin{figure}[h]
\begin{center}
\includegraphics[width=7cm]{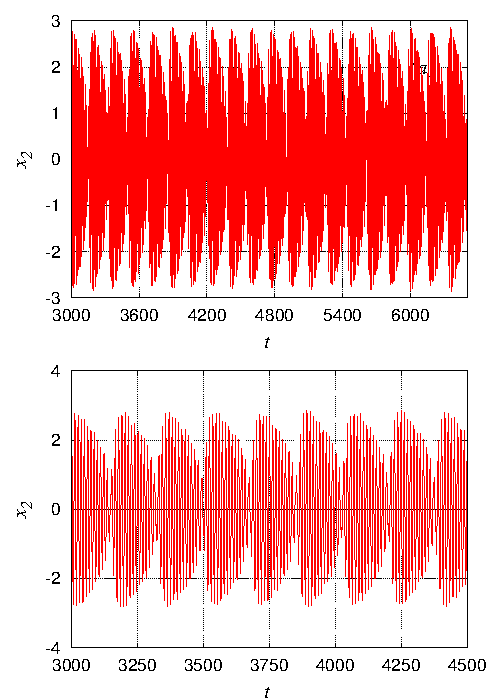}
\end{center}
\caption{\emph {Hyperchaotic Beats: Numerical Simulations (Event Driven Scheme)} : (a)The time series of the hyperchaotically modulated $x_2$ variable and (b) an expanded portion of the same for a short stretch of time.}
\label{Fig18}
\end{figure}
The presence of beats in any electronic circuit is explained in detail by \citet{rulkov96} as due to the incomplete synchronisation between self oscillations and the external forcing caused by a weak perturbation. The memristive MLC circuit can be considered to contain a resonant part, comprising of the $L-C$ combination and the memristor. In addition to this it has a nonlinear part whose role is solely taken by the memristor itself. The frequency of the circuit is determined by the resonant combination. The dissipative nature of the resonant part tries to destablise the stationary state of the circuit, while the non linearity, which is basically a time varying negative conductance is an active element and hence tries to restore the amplitude. Due to this interaction, a particular mode of periodic oscillations having a band of frequencies is preferred. Due to the existence of a band of resonant frequencies, the characteristic frequencies of this mode will change with a small variation of parameters of the circuit. An external periodic force applied to the circuit will modulate the $\it{environment}$ in which the intrinsic mode of oscillations lives. As a result of this modulation, the non linearity and dissipation will again tend to change the intrinsic mode of oscillation. If the perturbations introduced by the external periodic force is insufficient for proper change of intrinsic mode, then the circuit will be in and out of synchronisation intermittently and will demonstrate a partial collapse and revival of amplitudes of the circuit variables which we call as beats. While the presence of beats may be attributed due to this interaction between the self oscillations of the circuit and the external forcing, the nonsmooth nature of the circuit and the presence of grazing bifurcation of periodic orbits at the topological discontinuities are believed to endow the beats phenomenon with a hyper chaotic nature. Nevertheless, purely quasiperiodic beats may be observed when the initial conditions are so chosen that no grazing bifurcations occur.
\begin{figure}[h]
\begin{center}
\includegraphics[width=8cm]{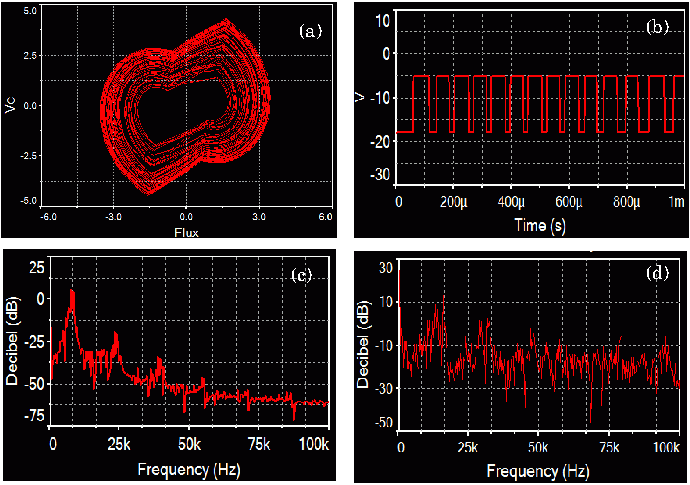}
\end{center}
\caption{ $\emph {Hyperchaotic Beats: Multisim Simulations}$ : The phase portrait in (a) the $(\phi-v_C)$ plane, (b) the variation in time of the memductance $W(\phi)$, (c) the power spectrum of the flux $\phi$ across the memristor and (d) the power spectrum of memristor switching showing it to be chaotically time varying resistor (CTVR).}
\label{Fig19}
\end{figure}
\begin{figure}[h]
\begin{center}
\includegraphics[width=8cm]{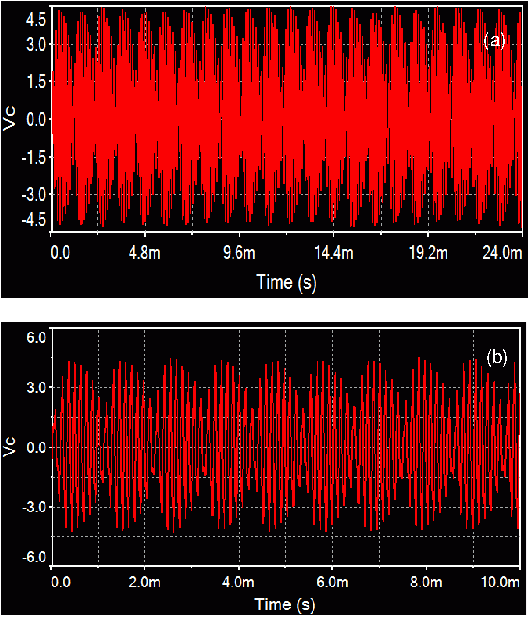}
\end{center}
\caption{\emph {Hyperchaotic Beats: Multisim Simulations } : (a)The time series of the hyperchaotically modulated $v_C$ variable and (b) an expanded portion of the same for a short stretch of time.}
\label{Fig20}
\end{figure}
To observe the hyper chaotic beats numerically we set the parameters as $\beta = 0.355263$, $a_1 = -0.697674$, $a_2 = -0.180433$. The eigen values of the unforced case (Sec. 3.1) for this choice of parameters are
\begin{eqnarray}
\sigma_1 = 0.0, |x_1| <1, \sigma_{2,3} = -0.526469 \pm i 0.570922,  \nonumber \\
\sigma_1 = 0.0, |x_2| >1, \sigma_{2,3} = -0.267848 \pm i 0.589594. 
\label{aia:eq23}
\end{eqnarray}
Here both the sets 
\begin{equation}
A_2 = \{ (x_1,x_2,x_3) |~~|x_1| < 1, x_2 = x_3 =  0 \}, \nonumber \\
\end{equation}
and
\begin{equation}
A_{1,3} = \{ (x_1,x_2,x_3) |~~|x_1| > 1, x_2 = x_3 = 0 \}. \nonumber \\
\end{equation}
are spirals or the system exhibits damped self oscillations with a natural frequency $\omega_o$ which is found to vary in the range  $0.156187 < \omega_o <     0.567041$. It is at this frequency that the memristor switching also occurs. This means $\omega_{mem} = \omega_o$. When an external driving force is applied, we find that the unstable damped oscillations stabilize to limit cycles and when the frequency of the external source is adjusted close to the memristor frequency, the phenomenon of beats occurs. Here in this circuit, when the normalized frequency of the external source is fixed as $\omega_{ext} = 0.508939$, and the normalized  amplitude of the driving force is kept as $f = 0.057846$, while all the other parameters are kept fixed as above,  hyper chaotic beats occur with the Lyapunov exponent values $(\lambda_1 = +0.125510$, $\lambda_2 = +0.016904,\lambda_3 = -0.107554 $ and $ \lambda_4 = 0.000000)$ as shown in Figs. (15) $\&$ (16). 
\section{Experimental Study}
We present here the results of the circuit simulation of hyper chaotic beats in the memristive MLC circuit using the Multisim model of the memristor mentioned earlier. The circuit implementation itself is shown in Fig. 3(b). The parameter values of the circuit  are fixed as $ L = 21 mH$, $R = 900 \Omega$, $C = 10.5nF$. For the memristor, the parameters are $ R_1 = 10 K\Omega$, $R_2 = 100 K\Omega$, $R_3 = 100 K\Omega$ and $C_3 = 2.2 nF$ for the integrator part, $R_4 = 10 K\Omega$ for the output of the window comparator, $R_5 = 1450 \Omega$ for the linear resistance and $R_6 = 1050 \Omega$, $R_7 = 2 K\Omega$ and $R_8 = 2 K\Omega$ for the negative conductance part. The reference voltages for the window comparator are fixed as $\pm 1V$. The frequency of the external sinusoidal forcing is fixed as $\nu_{ext} = 8.288 KHz$ and the amplitude is fixed as $F = 770 mV V_{pp}$. For these choice of parameters, we find hyper chaotic beats. The phase portrait in the $(\phi-v_c)$ plane and the time series plot of the memristor switching are shown in Figs. 19(a-b). The power spectrum  of the flux $\phi$ across the memristor as well as that of the memristor switching are shown in Fig. 19(c-d) respectively. The time series of the voltage across the capacitor for an extended range is shown in Fig. 20(a) and for a short span in Fig. 20(b). All these are similar to the numerical simulation results of the hyper chaotic beats phenomenon reported in Sec. 9.
\section{Conclusion}
In this study some of the questions that were considered at the begining have been addressed. Firstly the prototypical multisim model of the memristor, which we had proposed earlier and even designed a nonautonomous Chua's memrestive oscillator using it, has been successfully employed in designing a memristive MLC circuit. We have identified this memristive MLC circuit as a piecewise smooth system of second order with two discontinuous boundaries.  Grazing bifurcations which are a form of non-smooth bifurcations, have been found to occur in this circuit giving rise to chaotic, hyper chaotic and transient hyper chaotic dynamics. The bistability nature of the memristor, the interactions of the self oscillations with the external forcing and grazing bifurcations have given rise to hyper chaotic beats. These have been verified by numerical simulations after constructing proper ZDM and PDM mappings. Multisim modelling of the circuit has resulted in the verification of the concepts proposed and behaviors observed through numerical computations.

This work is important in that it has brought the investigation of analog circuits within the framework of non-smooth bifurcation theory. Earlier such bifurcations were identified and studied in detail in mechanical systems or power electronic circuits alone. As memristor has an inherent bistability, all memristive systems may supposed to be endowed with a non-smooth nature, and an analysis of these may help in unravelling their true dynamics, which would not be possible when one applies the theories and concepts developed for smooth systems. Already we have performed this analysis for the memristive Chua's oscillator and driven memristive Chua's oscillator. The results of this study will be published later. 

As further study, it is clear that the insights gained by the analysis of individual and networks of such oscillators may help in the building up of potential applications such as neuromorphic circuits and dense nonvolatile and dynamic memories. Further we feel that the identification of the hyper chaotic saddles and the control of transient phenomenon as well as use of hyper chaotic beats to physical and real time applications are some of the areas that can be further investigated. 

\section*{\label{Ack}Acknowledgements}

This work forms a part of a Department of Science and Technology (DST), Government of India, IRHPA project of ML and a DST Ramanna Fellowship awarded to him. ML has also been financially supported by a DAE Raja Ramanna Fellowship. The authors are thankful to Dr. K. Murali, Dr. K. Srinivasan and Mr. P. Megavarna Ezhilarasu for discussions and collaboration.

\section*{Appendix:~~ZDM and PDM Corrections for Grazing Bifurcations}{\nonumber}

In this section the conditions for grazing bifurcations to occur in a general second order non-smooth system having a single discontinuity boundary and the analytical expressions for the discontinuity mappings \cite{dib08} for transversal crossings of the periodic orbits are given. Using these, it is shown that, the ZDM correction and PDM correction given in Eqs. (\ref{aia:eq16}) $\&$ (\ref{aia:eq17}) respectively for the memristive MLC circuit can be obtained.

Let us consider a non-smooth system of $n^{th}$ order, having a single discontinuity surface defined by 
\begin{equation}
\Sigma  :=  \{ x \in \mathbb{D}: H(x) = 0 \}. \\
\label{aia:eqA.1}
\end{equation}
In the neighbourhood of this boundary the non-smooth system can be written locally as a system of smooth ODE's
\begin{equation}
\dot{x}(t) = 
	\begin{cases}
	F_1(x) & \text{if $H_1(x) \geq  0 $ }  \\
	F_2(x,) & \text{if $H_1(x) <  0 $ }.
	\end{cases}
\label{aia:eqA.2}
\end{equation}

For the dynamical system under consideration, the vector field $F_1(x)$ generates a flow function $\Phi_1$ while the vector field $F_2$ will generate a flow $\Phi_2$. Hence to detect this cross-over, we assume a scalar smooth function  $H(x)$, such that the zero set of this smooth scalar function defines the discontinuity surface $\Sigma _{12}$. A point $x = x^{\ast}$ is said to be a grazing point of a flow if the three conditions that follow are satisfied.
\begin{eqnarray}
\nonumber
H(x^\ast)    & = & 0, \\ \nonumber 
v(x^\ast)  & = & \frac{\partial}{\partial t} H(\Phi_1({x}^\ast,0) \\ \nonumber
&=&\frac{\partial H(x)}{\partial x}F_1(x^\ast), \\ \nonumber 
a({x}^\ast)  & = & \frac{\partial^2}{\partial t^2} H(\Phi_1({x}^\ast,0) \\ \nonumber
&=&\frac{\partial H(x)}{\partial x}\frac{\partial F_1(x^\ast)}{\partial x} F_1(x^\ast)\\
&:=& a^\ast > 0.
\label{aia:eqA.3}
\end{eqnarray}
All of these expressions are evaluated at the point $ x = x^\ast $. 

The first condition states that $x^\ast \in \Sigma $, whereas the second condition states that the flow is tangent to $\Sigma$ at $x^\ast$. The third condition gives the sense of the direction of the flows upon reaching the discontinuity surface and crossing it. A positive value, $a(x^\ast) > 0$, tells that the flow is convex to the boundary. However, a negative value for $a(x^\ast)<0$, tells that the flow is concave to the boundary.  In addition to these there is a transversality condition that is to be satisfied, namely 
\begin{equation}
\frac{\partial H_1(x)}{\partial x}F_1(x)\frac{\partial H_2(x)}{\partial x}F_2(x) \geq 0.
\label{aia:eqA.4}
\end{equation}  
This condition ensures that no sliding motion occurs along the discontinuity boundary.

The unique solution of the dynamical system can be obtained by monitoring the scalar function $H(x)$ during the numerical integration of the system equations and when $H(x)=0$, the so called Zero Discontinuity Map (ZDM) is applied. The zero discontinuity mapping takes into account both the changes in the state variables and the vector field before and after the discontinuity jump.

To obtain a periodic orbit we can reason out that it is possible to find at least one point, say $x_p^\ast$ which lies not on the discontinuity boundary but on a local transversal Poincar\'{e} surface $\Sigma_P$  such that $x_p^\ast \in \Sigma_P$. Here the local transversal Poincar\'{e} surface is defined using a scalar function $H_{\Sigma_P}(x_p^\ast)$ such that 
\begin{equation}
\Phi(x_p^\ast, T):= x_p^\ast H_{\Sigma_P}(x_p^\ast) = 0.
\label{aia:eqA.5}
\end{equation}
Solving these equations using Newton's method, the periodic orbits can be located. The periodic motions with grazing impacts on the discontinuity surface will then become a fixed point of this local transversal Poincar\'{e} mapping. These motions are sensitive to even small perturbations leading to eventual bifurcations. By constructing proper discontinuity maps, these grazing bifurcations can be simulated numerically.

Let $x^\ast$ be a regular grazing point of a second order piecewise-smooth system defined by Eq. (\ref{aia:eqA.2}). Then the Zero Discontinuity Map describing the trajectories of this system in a neighbourhood of the grazing trajectory has a 3/2-type singularity at the grazing point and is given as 
\begin{equation}
x \mapsto  ZDM(x)  = 	\begin{cases}
			x & \text{if $H_{min}(x) \geq  0 $ }\\
			x  + \delta & \text{if $H_{min}(x) < 0$ }.
	\end{cases}
\label{aia:eqA.6}
\end{equation}
 The correction $\delta$ is given as
\begin{equation}
\delta  =   2\sqrt{H_{min}} \sqrt{\frac{2}{(H_xF_1)_xF}} v(x) + O(|x|^2)
\label{aia:eqA.7}
\end{equation}
where  $v(x) = v_1 + v_2 + v_3 $ with $v_1, v_2,v_3 \in \mathbb{R}^n$ being each proportional to $x$ and are given by
\begin{eqnarray}
v_1 &=& -\{ -\frac{((H_xF_2)_x (F_1-\frac{2}{3}F_2))_x}{(H_xF_2)_xF}(F_2-F_1)_xF+  \nonumber \\
& &(F_{1,x}F_2-\frac{1}{3}F_{1,x}F_1-\frac{2}{3}F_{2,x}F_2)_xF \} \frac{H_x x}{(H_xF_1)_x F}, \nonumber \\
v_2 &=& (F_2-F_1)_x x, \nonumber \\
v_3 &=& -(F2-F1)_x F \frac{(H_x F_2)_x x}{(H_x F_2)_x F}, \nonumber \\
H_{min}(x) &=& \frac{\partial H}{\partial x}x+O(|x|^2). 
\label{aia:eqA.8}
\end{eqnarray}
 where $ O(|x|^2 $ is the correction term in the Taylor expansion of $H(x)$.

Let $x$ be a point in $\Pi_N$, the Poincar\'{e} section defined by
\begin{equation}
\Pi_N = \{ x \in \mathbb{D}: \frac{\partial H(x)}{\partial x}F_1(x) = 0 \}.\\
\label{aia:eqA.9}
\end{equation}
Then, sufficiently close to a regular grazing point satisfying Eq.(\ref{aia:eqA.3}) of a second order non-smooth system defined above, the Poincar\'{e}-section discontinuity map is given as
\begin{equation}
x  \mapsto PDM(x) =  	\begin{cases}
			x & \text{if $H(x) \geq  0 $ }\\
			x  + \gamma & \text{if $H(x) < 0$ }.
	\end{cases}	
\label{aia:eqA.10}
\end{equation}
where the correction $\gamma$ is given as
\begin{eqnarray}
\nonumber 
\gamma &=& v_{1z}(-H(x))^{\frac{3}{2}} + v_{2z}x(-H(x))^{\frac{1}{2}} + \\ 
& & v_{3z}H_xF_{2,x} x(-H(x))^{\frac{1}{2}} + O(x^2)
\label{aia:eqA.11}
\end{eqnarray}
where
\begin{eqnarray}
\nonumber
v_{1z}  &=& \frac{2}{(H_x F_{1,x} F)^{3/2}} \{\frac{1}{3} (F_{2,xx}-F_{1,xx})F^2 +  F_{2,x}F_{1,x}F\\ \nonumber
& &- \frac{1}{3}[F^2_{1,x}+2F_{2,x}]F- \frac{1}{H_x F_{2,x}F}(F_{2,x}-F_{1,x})\nonumber \\
         & & F[\frac{1}{3}H_xF_{2,xx}F^2+H_xF_{2,x}F_{1,x}F - \frac{2}{3}H_xF_{2,xx}F^2] \}, \nonumber \\
v_{2z} &=& \frac{2\sqrt{2}}{\sqrt{H_x F_{1,x} F}}(F_{2,x}-F_{1,x}), \nonumber \\
v_{3z} &=& \frac{2\sqrt{2}}{H_x F_{2,x}F \sqrt{H_xF_{1,x}F}}(F_{2,x}-F_{1,x})F. 
\label{aia:eqA.12}
\end{eqnarray}

The subscript $x$ denotes partial differentiation with respect to the variables $x = \{ x_1,x_2,x_3,x_4 \}$ and the subscript $z$ denotes the projection along z, the normal vector to $\Pi_N$. Further $F$ denotes that the vector fields are continuous at the discontinuity, that is 
$F_1(x^\ast) = F_2(x^\ast) = F(x^\ast)$.
Substituting $F_1$, $F_2$, $F$, $H(x)$ for the system defined by Eq. (\ref{aia:eqA.2}), the ZDM correction $\delta$ as well as the PDM correction $\gamma$ for the grazing orbit at the transversal intersection with the discontinuity surface $\Sigma_{1,2}$ can be obtained.

For the memristive MLC circuit, we define two scalar functions $H_1(x) = (x-x^\ast)$, $x^\ast = -1$ and $H_2(x)= (x-x^\ast)$, $x^\ast = +1$, such that their zero sets define the two discontinuity boundaries $\Sigma_{1,2}$ and $\Sigma_{2,3}$. 

When the conditions for transversal grazings given by Eq. (\ref{aia:eqA.3}) are satisfied at each of the discontinuity boundaries, then the ZDM correction $\delta$ given by Eq. (\ref{aia:eqA.7}) can be derived by applying Eq. (\ref{aia:eqA.2}) in Eq. (\ref{aia:eqA.8}). 

\begin{equation}
v_1  =  \left (	\begin{array}{c}
				-\frac{1}{3}(a_1-a_2)x_1\\
				\frac{1}{3}(a_1-a_2)(a_1-2(a_2+\beta ))x_1 +\frac{2(a_1-a_2)fx_1 \sin(\omega x_4)}{3x_3}\\
				\frac{1}{3}(a_1-a_2)\beta x_1\\	
				0 
				\end{array}
		\right ) .
\label{aia:eqA.13}
\end{equation}

\begin{equation}
v_2  =  \left (	\begin{array}{c}
				0\\
				(a_1-a_2)x_2\\
				0\\	
				0 
				\end{array}
		\right ) .
\label{aia:eqA.14}
\end{equation}

\begin{equation}
v_3  =  \left (	\begin{array}{c}
				0\\
				-(a_1-a_2)x_2\\
				0\\	
				0 
				\end{array}
		\right ) .
\label{aia:eqA.15}
\end{equation}

Substituting these, the ZDM correction term $\delta $ is given as

\begin{equation}
\small
\delta  =   \left ( \begin{array}{c}             						
				-\frac{2}{3}\sqrt{2}(a_1-a_2)x_1 \sqrt{\frac{x_1}{x_3}} \\
	                         \frac{2}{3}\sqrt{2}(a_1-a_2)(\frac{x_1}{x_3})^{3/2}{(a_1-2(a_2 + \beta))x_3 + 2fsin(\omega x_4)} \\
				\frac{2}{3}\sqrt{2}(a_1-a_2)\beta x_1 \sqrt{\frac{x_1}{x_3}} \\						
				0				 				    
				\end{array}
		\right).\\
\label{aia:eqA.16}
\end{equation}

Similarly the PDM correction can be derived by substituting Eq. (\ref{aia:eqA.2}) $\& $ (\ref{aia:eqA.12}), in  Eq. (\ref{aia:eqA.11}).

\begin{equation}
v_{1z}  =  \left (	\begin{array}{c}
				-\frac{4(1+a_1+\beta)}{3 \sqrt{3}}+\frac{4f\sin (\omega x_4)}{3\sqrt[3]{x_3}} \\
				\frac{2(a_1-a_2)(a_1+\beta)}{\sqrt{x_3}}+\frac{2(a_1-a_2)(a_1+\beta)}{\sqrt{x_3}}+ \\
				\frac{4(a_1^2+a_2+a_1\beta+\beta^2)}{3\sqrt{x_3}}+  \\
				\frac{6(a_1-a_2)(a_1+\beta)x_3+2(-2+a_1-3a_2-2\beta)f\sin(\omega x_4)}{3\sqrt[3]{x_3}} \\
				\frac{4\beta(1+a_1+\beta-\beta^2)}{3\sqrt{x_3}}+\frac{4\beta^2f\sin(\omega x_4)}{3\sqrt[3]{x_3}}\\	
				0 
				\end{array}
		\right ).\\
\label{aia:eqA.17}
\end{equation}
\begin{equation}
v_{2z}  =  \left (	\begin{array}{c}
				0\\
				\frac{2\sqrt{2}(a_1-a_2)}{\sqrt{x_3}}\\
				0\\	
				0 
				\end{array}
		\right ).\\
\label{aia:eqA.18}
\end{equation}
\begin{equation}
v_{3z}  =  \left (	\begin{array}{c}
				0\\
				\frac{2\sqrt{2}(a_1-a_2)}{\sqrt{x_3}}\\
				0\\	
				0 
				\end{array}
		\right ).\\
\label{aia:eqA.19}
\end{equation}

Making use of these the PDM correction term $ \gamma$ is given as
\begin{equation}
\gamma  =   \left ( \begin{array}{c}             						
				0 \\
	                        \frac{2(a_1-a_2)(a_1+\beta){x_1}^{3/2}}{\sqrt{x_3}}\\
				0\\						
				0				 				    
				\end{array}
		\right).\\
\label{aia:eqA.20}
\end{equation}

\bibliographystyle{apsrev4-1}
\bibliography{mem_mlc_updated.bib}
\end{document}